\documentclass[fleqn,usenatbib]{mnras}

\usepackage{newtxtext,newtxmath}

\usepackage[T1]{fontenc}

\DeclareRobustCommand{\VAN}[3]{#2}
\let\VANthebibliography\thebibliography
\def\thebibliography{\DeclareRobustCommand{\VAN}[3]{##3}\VANthebibliography}

\usepackage{graphicx}	
\usepackage{amsmath}	
\usepackage{courier}    
\title{New Methods of Cosmic Chronometer Identification Using Anomaly Detection}
\author[L. J. Hunt et al.]{Laura J. Hunt,$^{1,2,3}$
Kevin A. Pimbblet,$^{2,3}$
and David M. Benoit$^{2}$
\\
$^{1}$The School of Physical Sciences, The Open University, Walton Hall, Kents Hill, Milton Keynes, MK7 6AA, UK \\
$^{2}$E.A. Milne Centre for Astrophysics, Faculty of Science and Engineering, University of Hull, Cottingham Road, Kingston-upon-Hull HU6 7RX, UK \\
$^{3}$ Centre of Excellence for Data Science, AI, and Modelling (DAIM), University of Hull, Cottingham Road, Kingston-upon-Hull, HU6 7RX, UK
}

\date{Accepted XXX. Received YYY; in original form ZZZ}

\pubyear{2026}

\begin{document}
\label{firstpage}
\pagerange{\pageref{firstpage}--\pageref{lastpage}}
\maketitle

\begin{abstract}
We present a machine learning based pipeline to flag cosmic chronometer candidates based on SDSS imaging for future spectroscopic observation so their CC status can be confirmed. Our method utilises outlier detection with a convolutional autoencoder concurrently with a K-means clustering classifier to identify CC candidates between $0.03<z<0.34$ from the Galaxy and Mass Assembly (GAMA) survey with an accuracy, precision and recall of 0.819, 0.816, and 0.824 respectively. This method takes advantage of the difference between CCs and non-CCs when observed in the $u-$band and with the $u-g$ and $g-r$ colours. We evaluate the performance of the pipeline by assessing the misclassified samples, which we find to be intermediate-type galaxies that have properties similar to that of the correctly classified CC and non-CC samples. Finally, we compare the results of the $H_0$ calculations, calculated with the differential age–redshift method, with parameter estimation performed using MCMC sampling. When using ML classified CCs with and without contamination of misclassified non-CCs, we find that the effect of contamination increases the uncertainty on $H_0$.

\end{abstract}

\begin{keywords}
methods: data analysis -- galaxies: evolution -- cosmology: observations -- cosmology: distance scale -- dark energy
\end{keywords}

\section{Introduction} \label{intro}
The tension between early and late Universe measurements of the Hubble constant currently sits at a $\sim4-6\sigma$ difference \citep[for reviews:][]{2014PDU.....5..307V, 2021CQGra..38o3001D, 2021ApJ...919...16F, 2021A&ARv..29....9S} which makes the topic of Universal expansion one of the most important issues in current cosmology. One method that may resolve the tension by providing a late Universe, model-independent measurement without needing to calibrate the distance ladder is based on the differential age-redshift relation \citep{2002ApJ...573...37J}. This method allows us to probe expansion directly between objects that have formed at the same time but have moved apart due to expansion.

\begin{equation}
    H(z) \equiv \frac{-1}{1+z}\frac{\Delta z}{\Delta t}
\end{equation}

Where $H(z)$ is the Hubble parameter at an effective redshift between two objects separately by a difference of redshift $\Delta z$ and age $\Delta t$ \citep[e.g.][]{2014PDU.....5..307V, 2016JCAP...05..014M, 2019JCAP...03..043J, 2022JHEAp..36...27V, 2022ApJ...928L...4B, 2023ApJS..265...48J, 2025MNRAS.540.3135L}. This method relies on the idea of ``standard clocks'' or cosmic chronometers that have a uniform formation history that allows them to be used to measure the age of the Universe. The identification of cosmic chronometer (CC) candidates is contingent on ensuring that even minimal star formation is eliminated from any given sample. Consequently, they are generally massive and passively evolving elliptical galaxies that form their initial stellar population rapidly, using up their gas reservoirs in the process, leaving them quiescent with no subsequent star-formation events triggered by mergers. Galaxies such as these are therefore rare, as the evolution of typical galaxies features at least some merging events. 

There is extensive research on the identification of massive, passively evolving elliptical galaxies and evaluation of contamination caused by younger stellar populations and how this may bias measurements of spectra \citep[e.g.][]{2012JCAP...08..006M, 2016JCAP...05..014M, 2018ApJ...868...84M, 2022ApJ...927..164B, 2023ApJS..265...48J, 2023A&A...679A..96T}. \cite{2018ApJ...868...84M} describes a detailed pipeline to select cosmic chronometer candidates by examining the effect of different fractions of young stellar components to their older counterparts. Their method describes five selection cuts to minimise the contamination with the addition of full spectral fitting analysis using traditional spectral fitting models such as VESPA \citep{2007MNRAS.381.1252T} or MOPED \citep{2000MNRAS.317..965H} before finally labelling a candidate as a bona fide cosmic chronometer. Though this method is comprehensive and robust, it requires extensive amounts of spectroscopic observations in order to identify even a small handful of CCs. For example, in the Large Early Galaxy Astrophysics Census (LEGA-C) \citep{2018ApJS..239...27S, 2016ApJS..223...29V} survey there are 350 identified CCs across $0.6<z<1$ in the COSMOS field identified from 1622 sources \cite{2022ApJ...927..164B}. \cite{2023A&A...679A..96T} identifies 49 CCs from 2087 galaxies across $1 < z < 1.5$ in the fourth data release of the VANDELS survey \citep{2018MNRAS.479...25M, 2020MNRAS.492.2128P, 2021A&A...647A.150G}. Larger samples of CCs have been assembled by using less strict selection criteria \citep{2016JCAP...05..014M, 2025MNRAS.540.3135L} however this risks introducing larger uncertainties from potentially mixed stellar populations. 

Ideally, to use CCs in the differential age-redshift method, one would require a wide redshift coverage to overcome the effects of age uncertainty however this requires deep spectroscopic surveys which are costly and time-consuming. Therefore, an alternative method of quickly identifying CC candidates by combining machine learning (ML) techniques with galaxy images could alleviate this issue by providing a fast ``pre-classification'' to flag potential CC candidates as targets for future spectroscopic observation. ML is a widely used tool in Astronomy and when used responsibly can aid in tasks previously hindered by issues relating to the processing of copious amounts of data produced by modern-day astronomical surveys. Classification algorithms are a common example of how ML is used in Astronomy and is successfully applied to the classification of galaxies based on various properties using both images and spectra \citep[e.g.][]{2024MNRAS.527.1163W}. Morphological classification is successfully achieved using supervised and unsupervised ML techniques, with a wide variety of algorithms that have been developed in the last few decades. Recent work using convolutional neural networks (CNNs) successfully classifies galaxies based on their structural morphology \citep[e.g.][]{2020MNRAS.491.1554W, 2022KosNT..28a...3V, 2022KosNT..28e..27K}. CNNs are a supervised ML model that learns which structures in an image are associated with a given label. Once the relationship between the images and labels is learned, they are then able to predict the classification of new images. However, because they are very structure orientated, they may not be able to differentiate CCs from normal elliptical galaxies. 

Additionally, similar examples of supervised ML based classification requires large, correctly labelled, balanced datasets for training. This means each label has an equal number of samples in the dataset, such that each label is equally common in the dataset. This is a major issue for astronomical applications of ML because assembling completely balanced datasets is not always possible if a particular object is rare. Unsupervised ML methods do not require samples to be labelled which allows algorithms to determine their own interpretation of the relationships present in a dataset. Outlier (or ``anomaly") detection is an application of unsupervised ML that uses dataset imbalance as a tool to identify anomalous objects. It relies on finding intrinsic relationships in observable data in order to flag samples that do not fit these trends \citep{2021inas.book..197D}. \cite{2017MNRAS.465.4530B} successfully applies an unsupervised Random Forest to SDSS spectra to identify anomalous galaxies, finding a significant fraction of the 400 most outlying samples are severe outliers on the BPT diagram, extremely H$\delta$ strong or H$\alpha$ weak. New large surveys such as Euclid even incorporate bespoke unsupervised algorithms for classification and outlier detection in order to process the vast amounts of high-dimensional data they produce \citep{2023A&A...671A..99E, 2025arXiv250315312E}. 

The relationship that we aim to capitalise on is the link between photometry and spectroscopy to make identification through imaging alone a possibility and therefore allow us to quickly probe larger regions on the sky in a more time efficient manner before potentially revisiting promising targets for follow-up spectroscopic observation and analysis. Photometric images can be successfully used to generate full spectra and recover various properties including colour bimodality, mean stellar age and velocity dispersion by linking morphological features and photometric colour with the physical quantities derived from spectra using a ML based conditional diffusion model \citep{2024ApJ...977..131D}. Additionally, recent Euclid data processing efforts find that different colours can be used to identify quiescent galaxies from star-forming galaxies using an ML based pipeline \citep{2023A&A...671A..99E}.

Autoencoders (AEs) are an unsupervised ML technique that is based on the concept of dimensionality reduction \citep{SANGER1989459}. The basic architecture includes an encoder made up of fully connected layers that take input data and reduces its dimensions to a bottleneck or latent space. A decoder then takes data that has passed through the bottleneck and reconstructs it, attempting to match the input as closely as possible. With this framework, the autoencoder can learn which features of the input data must be retained in the latent space in order to recreate the input. This means they are commonly used for dimensionality reduction, data compression and feature extraction. There is extensive research surrounding AEs for applications, including reconstructing galaxy images and spectra in Astronomy. In addition, different frameworks of AEs possess different utility. For example, typical AEs with a latent space represented by minimal dimensions are very successful at reproducing 1-dimensional data such as spectra. AEs can be used for outlier detection of galaxy spectra though the addition of normalising flows is included to separate anomalous data from normal data \citep{2023AJ....166...75L, 2023ApJ...956L...6L}. 

Convolutional autoencoders (CAEs) work on the same principles but function differently as they incorporate convolutional layers that are common for image based applications \citep{masci2011stacked, xie2012image}. The addition of convolutional layers instead of fully connected layers means the autoencoder is able to capture spatial hierarchies in images by using filters that pass over the input images. This makes CAEs more effective at preserving local structures in images compared to standard autoencoders. Thus, while both serve similar purposes, CAEs are better suited for tasks involving image compression, denoising, or feature extraction. 

\cite{2019PASP..131j8011R} applies a CAE on radio-astronomical images from Radio Galaxy Zoo (RGZ) in order to create self-organising maps (SOMs) from the latent vectors of the RGZ images which a K-means clustering algorithm classifies based on the radio-morphology latent features. \cite{2023PASP..135h4101S} similarly uses a CAE to generate latent features of $g-$, $r-$ and $i-$band images that they then use to create a database of latent features in which they apply similar image retrieval to classify galaxies. However, they find that limitations of the CAE include the colour, as redder LTGs are often misclassified as ETGs and vice versa. They suggest this may be due to the model not picking up on the finer detail of spiral arms, instead relying on the overall colour of images. They suggest this could stem from their approach that involves taking $g-$, $r-$ and $i-$band images and creating the equivalent of a full-colour ``RGB'' image to input into a one-channel CAE rather than training a three-channel CAE or three completely separate models. Similar methods are applied to Galaxy Zoo 2 data with the aim of improving accuracy by exploiting both professionally classified galaxy images with amateur classifications created using citizen science. Other recent applications of CAEs are used to classify supernovae remnants \citep{2024A&A...690A.203B} and strong lenses \citep{2020MNRAS.494.3750C}. CAEs have also been successfully used to denoise galaxy images \citep{2023MNRAS.521.6318B, 2025A&A...693A.141F} and deconvolve point spread functions \citep{2024PhRvD.110j3030S}.

In this work, we use a CAE-derived anomaly metric, combined with observed-frame colours, to define an image-based pre-selection of spectroscopically selected CC-like galaxies. This is with the aim of determining if it is possible to identify CCs with imaging and photometric observation which has not been established before. Our goal is to capitalise on the ability of convolutional layers to identify important structure across five different photometric bands in order to identify CCs based on their limited morphology in different colours. We describe our sample selection in Paper I \citep{2026MNRAS.tmp.1629H} but include an overview of the image preparation needed to create our training sample in section~\ref{data}. We provide a full description of our CAE and the method of CC classification in section~\ref{method}. The results of the CAE model and outlier identification can be found in section~\ref{cae_results} and section~\ref{class_results} respectively whilst the comparison between the derived values of $H_0$ using different samples of CCs can be found in section~\ref{calc_h0}. Finally, we give a discussion of these results and comparison to the literature in section \ref{disc}. For illustrative purposes only we use Planck18 \citep{2020A&A...641A...6P} values for $H_0=68$~kms$^{-1}$Mpc$^{-1}$, $\Omega_\text{m}=0.31$ and $\Omega_{\Lambda}=0.69$ in figures showing comparisons of $H_0$ with the literature.

\section{Training Data} \label{data}
To assemble a dataset for the training and testing of our convolutional autoencoder (CAE), we first create a sample of classified CCs and non-CCs. To do this we use a selection of data from the Galaxy and Mass Assembly (GAMA) survey \citep{2009A&G....50e..12D, 2015MNRAS.452.2087L} which includes photometry, spectra and stellar mass estimates combined with stellar velocity dispersions from SDSS DR17 \citep{2000AJ....120.1579Y, 2011AJ....142...72E, 2012AJ....144..144B, 2022ApJS..259...35A}. The process of assembling our final sample begins with making a number of selection cuts based on those described by \cite{2018ApJ...868...84M}. This process leaves us with a dataset of 231 high-purity CC candidates and 8,290 confirmed non-CC galaxies which we cross-match with SDSS imagery to assemble the final galaxy image dataset which is used for training the CAE. For a detailed description of the selection process and an analysis of the stellar populations, please see paper I \citep{2026MNRAS.tmp.1629H} in which we identify the CC candidates from the GAMA survey between $0.03<z<0.34$. 

\begin{figure}
    \centering
    \includegraphics[width=\linewidth]{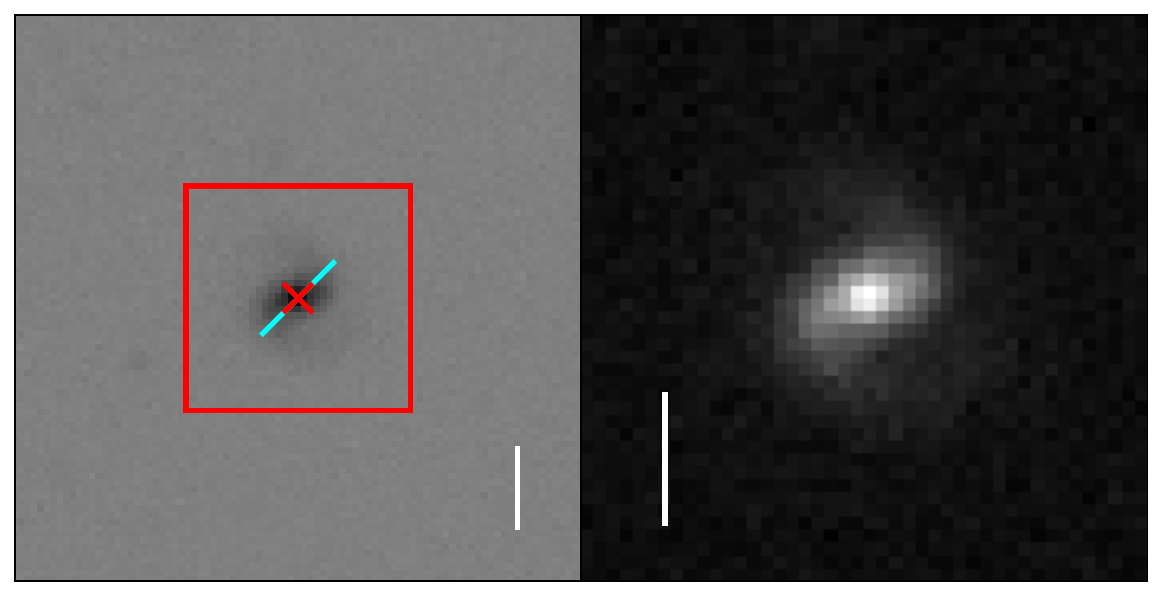}
    \caption{The \textit{left panel} shows an example of the cropping procedure for an SDSS galaxy in the $r$-band. The example is pre-cropped for demonstration purposes but in reality is a 1489x2048 SDSS frame. The target centre is marked with a red cross based on the galaxy's RA and Dec. The cyan line marks 2 times the length of the $r-$band Petrosian radius according to SDSS measurements. The red box enclosing the galaxies marks the dimensions that we crop to which correspond to 4 times the length of the major axis. The \textit{right panel} shows the result of the cropping procedure in which the white bars correspond to 10 kpc for demonstration purposes.}
    \label{fig:cropping}
\end{figure}

After cross-matching our sample with SDSS DR17 $ugriz$ frames to compile our image based dataset for CAE training, we are left with only 199 CCs with matching SDSS imaging. For each galaxy in the sample, we crop its corresponding $ugriz$ frames based on the $r-$band Petrosian radius. Fig.~\ref{fig:cropping} shows an example of the cropping procedure in which images are cropped to 4 times the $r-$band petrosian radius around the galaxy's target RA and Dec. Once all 5 images for each galaxy are cropped, we resize\footnote{By converting to a \texttt{PIL} (Python Imaging Library) \texttt{Image} and utilising the \texttt{PIL} resize function} to 128x128 pixels and then normalise\footnote{pytorch.transforms.pil\_to\_tensor} each image separately to $(0, 1)$ as the CAE requires images with standardised dimensions. 

We ensure the 199 CCs are removed from the CAE training set and select a further random 199 non-CCs to set aside from the remaining sample in order to compare CC versus non-CC performance without biasing the results by showing the CAE images it has been trained on. Finally, we take a train-test split of 80\% and 20\% respectively on the remaining sample to use during the training and testing phases. 

\section{Method} \label{method}
\subsection{Convolutional Autoencoder} \label{cae}
We use a convolutional autoencoder (CAE) which we implement with \texttt{PyTorch} \citep{2019arXiv191201703P}. Similarly to a traditional AEs, convolutional autoencoders consist of an encoder, bottleneck and decoder structure however, with the addition of convolutional and deconvolutional layers in the encoder and decoder. This means each layer is connected locally to the previous layer with a receptive field rather than in a fully-connected manner like in a traditional AE which allows us to retain the hierarchical structure within portions of the images. When an image is passed through a convolutional layer, a filter (or multiple filters) passes over each pixel, performing a convolution of the pixels under the filter. The weighted sum of each region (surrounding each pixel) is the result for one single output node which is defined by \texttt{PyTorch} as:

\begin{equation}
    \text{out}(N_i,C_{\text{out}_j})=\text{bias}(C_{\text{out}_j})+\sum_{k=0}^{C_{\text{in}}-1}\text{weight}(C_{\text{out}_j},k)*\text{input}(N_i,k)
\end{equation}

for an input of $[N,C_{\text{in}},H,W]$ and output $[N,C_{\text{out}},H_{\text{out}},W_{\text{out}}]$ where $N$ is the batch size, $C$ is the number of channels and $H$ and $W$ is the height and width in pixels respectively. During training, the CAE updates the weights through backpropagation to minimise a loss metric as much as possible. In each training epoch, it cycles through the entire training set in batches. For each batch, it calculates the reconstruction loss between the input and outputs, backpropagates the losses to the model parameters then updates the gradients of the model according to a chosen optimiser function. 

\subsubsection{Architecture} \label{method:ae}
\begin{figure*}
    \centering
    \includegraphics[width=\linewidth]{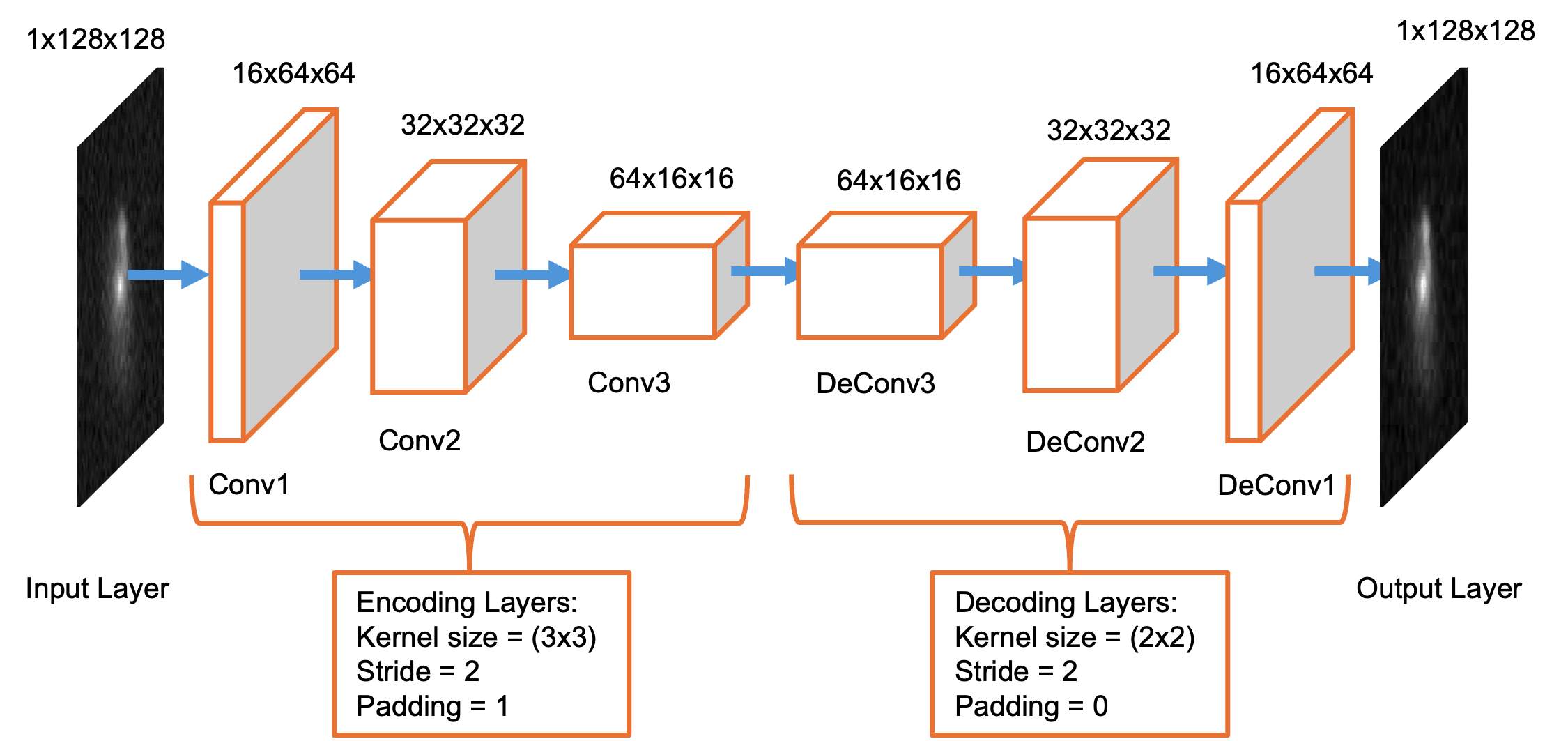}
    \caption{Our CAE framework includes an input layer, three 2D convolutional layers in the encoder, three 2D transposed convolutional layers in the decoder and an output layer.}
    \label{fig:ae_arch}
\end{figure*}

Our CAE architecture consists of an input layer, three convolutional layers in the encoder, three transposed convolutional (deconvolutional) layers in the decoder and an output layer, shown in Fig.~\ref{fig:ae_arch}. Our encoding layers take a 128x128 image, normalised between 0 and 1, through the input layer which is passed to the first convolutional layer\footnote{\texttt{PyTorch} nn.Conv2d}. We find three layers in the encoder and three layers in the decoder provide sufficient depth for the network to learn increasingly abstract representations of the image, while avoiding unnecessary architectural complexity for the relatively small input size and reconstruction task. All convolutional layers in the decoder rely on a kernel size of 2x2, a stride of 2, and padding of 0 to upsample the images rather than using a true deconvolution. The convolutional layers in the encoder use 3x3, a stride of 2 and padding of 1 to reduce dimensionality rather than using pooling layers. This is to retain as much information per pixel as possible, as the filters capture the spatial hierarchy of the image features. Pooling layers do not necessarily provide any additional learning to the network as they only perform a mathematical operation over the image to reduce its dimensions \citep{2014arXiv1412.6806S}. Instead, by using kernel size, stride and padding to reduce the height and width of the input image we create more depth with additional feature maps that highlight different learnable features. Additionally, typical autoencoders feature fully-connected layers between the encoder and decoder to produce a smaller latent space. This is beneficial for low-dimensional input data and high levels of dimensionality compression, but can cause important hierarchical features to be lost \citep{2014arXiv1411.4038L}. For our work, we are only interested in whether the CAE is able to reconstruct CC images as accurately as non-CC images. Therefore, we do not require a small bottleneck because we are trying to find a balance between high-dimensionality to retain as much information within the images as possible to flag possible outliers (CCs).

For each SDSS filter, we train a separate model referred to as the $u-$model, $g-$model, $r-$model, $i-$model and $z-$model herein. Each trained model has the same architecture but different weights due to being trained on different bands. We then train each model separately over 20 epochs using the Adam\footnote{\texttt{PyTorch} torch.optim.Adam} optimiser with a learning rate set to 0.001. For all layers except the final output layer in the decoder, we use the rectified linear unit (ReLU)\footnote{\texttt{PyTorch} F.relu} activation function. We use ReLU to ensure pixel values remain greater than zero without limiting their positive values. For the final output layer, we use the sigmoid\footnote{\texttt{PyTorch} F.sigmoid} activation function to ensure all reconstructed pixel values are normalised between 0 and 1. \texttt{PyTorch} applies both activation functions element-wise for batches of inputted tensors. 

When choosing a loss function, we compare the structural similarity index measure\footnote{\texttt{PyTorch} ignite.metrics.SSIM()} (SSIM), mean squared error\footnote{\texttt{PyTorch} nn.MSELoss()} (MSE) and L1\footnote{\texttt{PyTorch} nn.L1Loss()} losses. We find L1 performs marginally better for our training set so we continue to use L1 for training. L1 is less sensitive to outlying data than SSIM and MSE \citep{2015arXiv151108861Z} which is beneficial for the training phase as all of the training set are non-CC. In theory, this means that as the models are training, if they encounter a galaxy that they cannot reconstruct well (i.e. a higher reconstruction loss is calculated), L1 minimises the loss whereas a loss function such as MSE exaggerates it. When the weights are updated through backpropagation, the smaller L1 loss has less of an effect than a higher MSE loss value. Therefore, we select L1 loss because of its robustness to large reconstruction losses. This behaviour is beneficial to us because we want our network to be ``picky'' and perform worse with data that does not resemble our training set (i.e. it should be unable to reconstruct CC images if it has only been trained on non-CC images). 

\subsection{Classification Algorithm}
To perform the classification of CC or non-CC, we use a K-means clustering algorithm. K-means clustering is a geometric based unsupervised ML algorithm that categorises unlabelled data into groups based on the locations of centroids \citep{2015AnDS....2..165X}. K-means aims to minimise the within-cluster sum of squares (inertia) by finding the optimal number of centroids $k$. Typically, this is achieved by iteratively adding more centroids (increasing the value of $k$) and assigning data to the nearest centroid based on the shortest calculated Euclidean distance. The ideal number of clusters is then determined using a metric such as the silhouette score and an elbow graph. However, we use K-means as a binary classifier by assigning data to only two centroids (CC or non-CC). By varying the input features, we aim to find the ideal features that maximise the separation between the two clusters in order to find which properties of the CCs are needed to differentiate them from normal non-CC galaxies. We utilise K-means clustering using the Scikit-Learn K-means\footnote{sklearn.cluster.KMeans} algorithm. We first scale our input data using a standard scaler\footnote{sklearn.preprocessing.StandardScaler} to avoid biasing the derived relationships. 

\section{CAE Model Results} \label{cae_results}
We train the models using one compute node with 28 cores and 128GB of RAM on the Viper high-performance computer (HPC), which takes approximately 20 minutes for 20 epochs. Once trained, we use the 398 galaxies (199 CCs and 199 non-CCs) that are set aside from the training and validation sets. This ensures all analysis is completed with images that the models have not encountered whilst training. It is important to note that the CAE is not is not provided labels for any of the training or testing data because it is entirely unsupervised, though we do use the labels in order to measure performance after training and testing. Whilst evaluating the models, there are two important aspects we must consider in order to produce a pipeline for differentiating between CC and non-CC images:

\begin{enumerate}
    \item Choice of evaluation metric: which metrics of evaluation capture the difference between the CC and non-CC images by maximising the loss the most? 
    \item Choice of filter/band: in which filter do the CCs and non-CCs look the most different? With that in mind, which model trained on which filter performs best at differentiating the two samples?
\end{enumerate}

Firstly, we investigate the choice of reconstruction loss across the test sample by comparing a number of different metrics. We use binary cross entropy\footnote{\texttt{PyTorch} torch.nn.BCELoss} (BCE), structural similarity score\footnote{\texttt{scikit-image} skimage.metrics.structural\_similarity} (SSIM), peak signal to noise ratio \footnote{\texttt{scikit-image} skimage.metrics.peak\_signal\_noise\_ratio} (PSNR), mean squared error (MSE)\footnote{\texttt{scikit-image} skimage.metrics.mean\_squared\_error} and L1 loss which is functionally the same as mean absolute error when using the default parameters in the \texttt{PyTorch} function torch.nn.L1Loss.

BCE is a common metric used for reconstruction loss, during training and outside of training. It treats each pixel like a binary classification which is calculated by:

\begin{equation}
    \text{BCE} = -\frac{1}{N} \sum_{i=1}^N {[y_i\cdot\text{log}(\hat{y}_i)+(1-y_i)\cdot\text{log}(1-\hat{y}_i)]}
\end{equation} \label{eq:bce}

where $\hat{y}_i$ is the predicted image and $y_i$ is the input image. SSIM is defined as:

\begin{equation}
    \text{SSIM}=\frac{(2\mu_y \mu_{\hat{y}}+C_1)(2\sigma_{y\hat{y}}+C_2)}{(\mu_y^2+\mu_{\hat{y}}^2+C_1)(\sigma_y^2+\sigma_{\hat{y}}^2+C_2)}
\end{equation}

where $\mu_y$ and $\mu_{\hat{y}}$ are the mean brightness of the input and output images, $\sigma_y^2$ and $\sigma_{\hat{y}}^2$ are the image variances, $\sigma_{y\hat{y}}$ is the covariance between the two images and $C_1$ and $C_2$ are two stability constants that are equal to $C_1=(K_1 \cdot L)^2$ and $C_2=(K_2 \cdot L)^2$. Where $L$ is the data range ($L=1$ for normalised images), $K_1=0.01$ and $K_2=0.03$ as this is the default for \texttt{scikit-image} SSIM.

PSNR is typically used to compare a noise-free monochrome image with a noisy version, but is only used for comparing reconstructed images to their inputs rather than as a loss function during training. Higher PSNR values are generally correlated with higher quality reconstruction, which is defined by:

\begin{equation}
    \text{PSNR}=20\text{log}_{10} \left( \frac{MAX_y}{\sqrt{MSE}} \right)
\end{equation} \label{eq:psnr}

where $y$ is the input image, $MAX_y$ is the maximum pixel value in the input image and $MSE$ is the mean squared error defined by:

\begin{equation}
    \text{MSE}=\frac{1}{n}\sum^{n}_{i=1}{|y_i-\hat{y}_i|^2}
\end{equation}

where $\hat{y}$ is the reconstructed image. 

L1 loss is similar to MSE as it also measures the errors between actual and predicted values, as shown:

\begin{equation} \label{eq:mae}
    \text{L1} = \frac{1}{n} \sum_{i=1}^{n} |y_{i} - \hat{y}_{i}|  
\end{equation}

where $\hat{y_{i}}$ is the predicted value, $y_{i}$ is the corresponding true value and $n$ is the number of data points. Similarly to MSE, the closer to zero, the more accurate the predictions are. We choose to use both of these metrics because even though they are similar, they show slight nuances about the prediction ability of the CAE for example the L1 metric is less sensitive to large errors in prediction whereas MSE is able to penalise this more. 

\begin{table} 
    \caption{We find the mean absolute difference between the CC and non-CC losses to evaluate which metric shows the largest difference. L1 and MSE $u-$model losses show the largest difference with MSE performing marginally better.} \label{tab:all_loss_diff} 
    \begin{tabular}{c|ccccc} 
        Band & L1 & MSE & SSIM & BCE & PSNR \\ 
        \hline 
        $u$ & 0.2551 & 0.2554 & 0.1197 & 0.0161 & 0.1197\\ 
        $g$ & 0.1627 & 0.0982 & 0.1106 & 0.0023 & 0.1106\\ 
        $r$ & 0.1418 & 0.0905 & 0.0813 & 0.0013 & 0.0813\\ 
        $i$ & 0.1819 & 0.1329 & 0.0812 & 0.0013 & 0.0812\\ 
        $z$ & 0.1449 & 0.0768 & 0.1273 & 0.0039 & 0.1273\\ 
        \end{tabular} 
\end{table} 

\begin{figure}
    \centering
    \includegraphics[width=\linewidth]{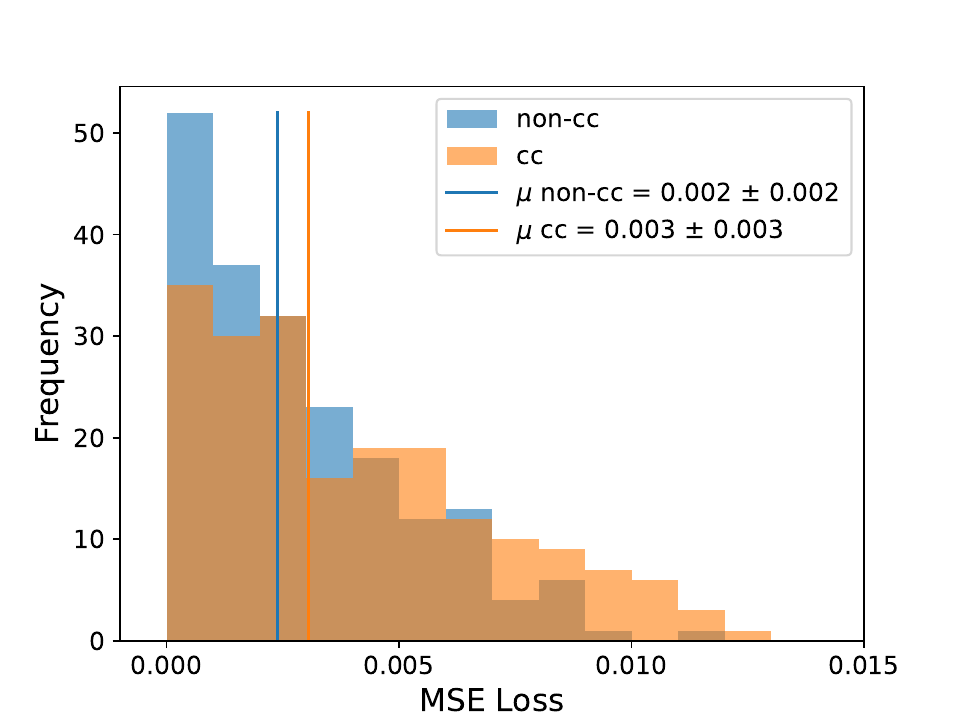}
    \caption{Histogram to show the MSE loss. There is a slight difference in mean $u-$model MSE values across the CC and non-CC samples but there is no separation between the distributions themselves.}
    \label{fig:u_mse_hist}
\end{figure}

To evaluate the performance of these metrics, we normalise the average loss difference for each metric in order to compare the difference between the CC and non-CC values across the 5 models which is shown in Table~\ref{tab:all_loss_diff}. L1 and MSE $u-$model losses show the largest difference with MSE performing marginally better. We then evaluate each model using the MSE loss by calculating the difference in the mean non-CC and CC MSE losses across the 5 models in Table~\ref{tab:Normal_mse}. The $u-$model shows a significantly larger difference than any of the other models with a difference of 0.000661. We show the distribution of $u-$model MSE losses for the CC and non-CC samples in Fig.~\ref{fig:u_mse_hist}, along with the mean values of each sample. Though there is a difference between the means, there is still a significant overlap between the two samples which suggests a more sophisticated method of classification needs to be employed to tease apart the two samples.

\begin{table} 
    \caption{The average MSE loss for CCs and non-CCs when reconstructed by the 5 models. The $u-$model shows the largest difference with the CC samples having slightly worse MSEs on average. } 
    \begin{tabular}{c|cccc} 
        Model & Non-CC & CC & Both & Difference \\ 
        \hline 
        $u$ & 0.002395 & 0.003056 & 0.002680 & 0.000661 \\ 
        $g$ & 0.000088 & 0.000086 & 0.000088 & -0.000002 \\ 
        $r$ & 0.000055 & 0.000047 & 0.000051 & -0.000008 \\ 
        $i$ & 0.000057 & 0.000046 & 0.000051 & -0.000011 \\ 
        $z$ & 0.000185 & 0.000129 & 0.000155 & -0.000056 \\ 
    \end{tabular} 
    \label{tab:Normal_mse}  
\end{table} 

\begin{figure*}
    \centering
    \includegraphics[width=\linewidth]{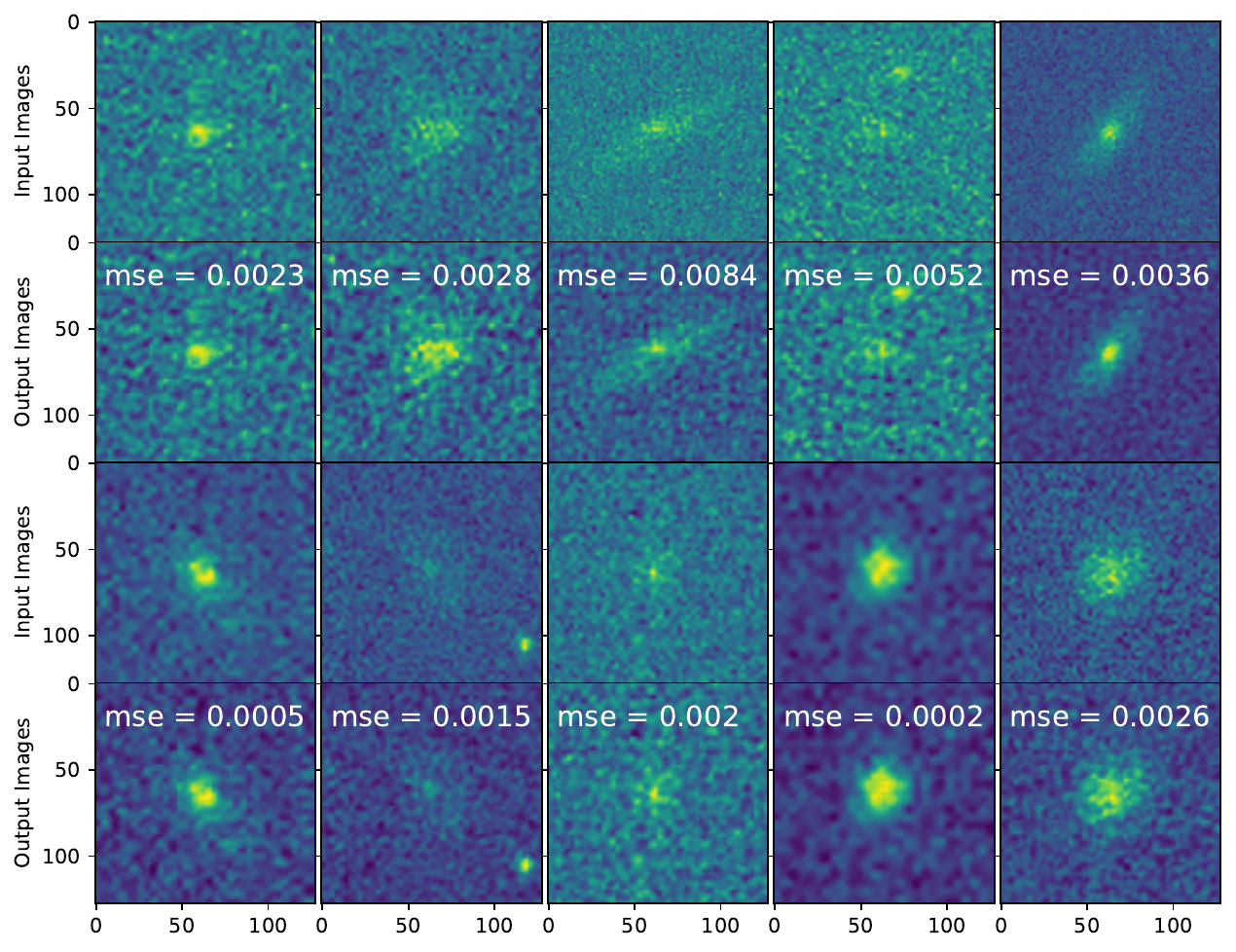}
    \caption{Examples of non-CCs input and reconstructed images shown in the $u-$band with MSE loss for each image overlaid.}
    \label{fig:norm_nonccs}
\end{figure*}

\begin{figure*}
    \centering
    \includegraphics[width=\linewidth]{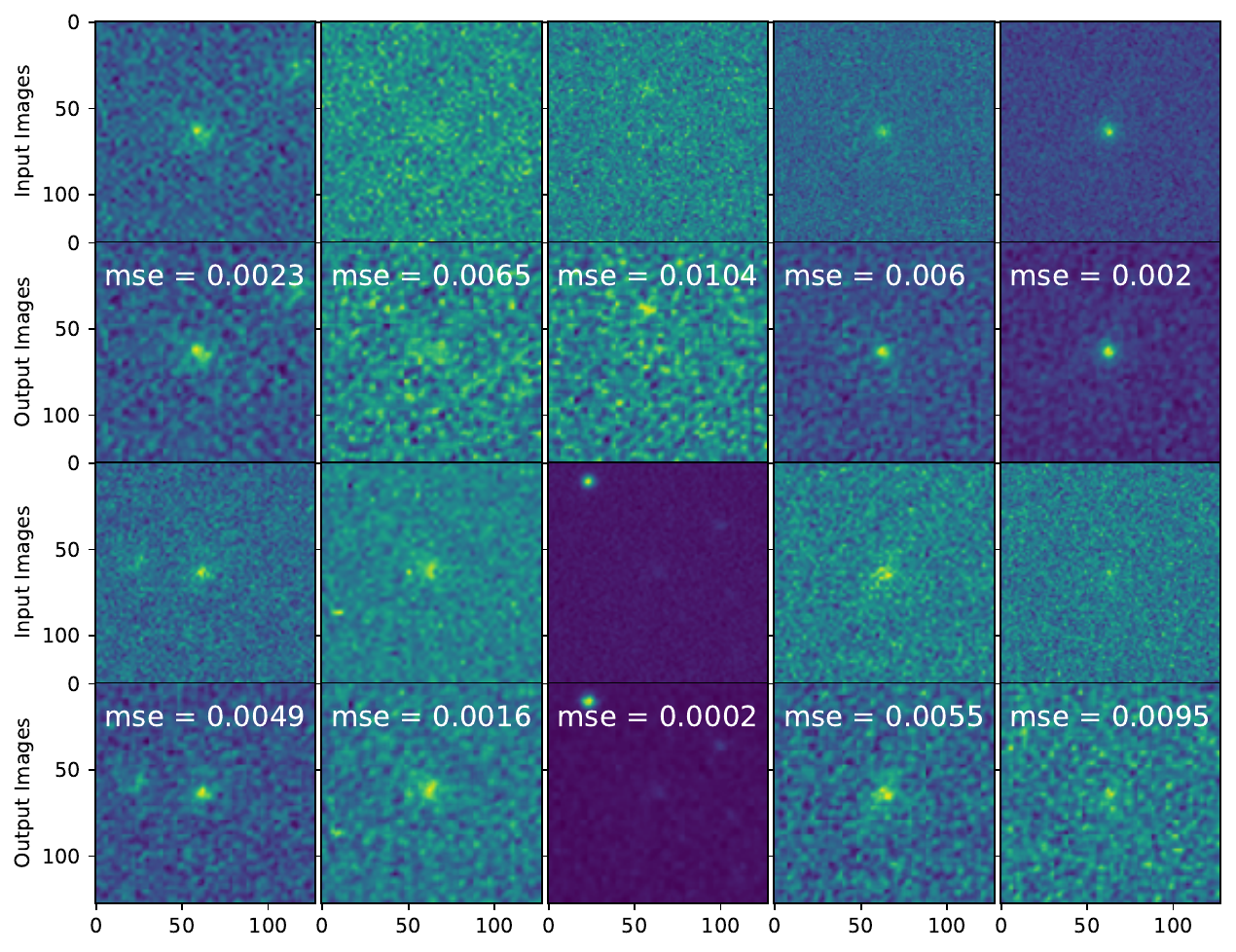}
    \caption{Examples of CCs input and reconstructed images shown in the $u-$band with MSE loss for each image overlaid.}
    \label{fig:norm_ccs}
\end{figure*}

Finally, Fig.\ref{fig:norm_nonccs} and Fig.\ref{fig:norm_ccs} shows five examples of non-CCs and CCs respectively alongside their reconstructed images. All images appear to be reconstructed well with the exception of some slight pixelation in the reconstructions. The $u-$band images appear to be noisy but this is likely due to the overall brightness being lower in the bluer wavelengths which makes the background stand out more. The CC $u-$band images Fig.\ref{fig:norm_ccs} show that the network could be trying to recreate nonexistent structure as there appears to be a brighter centre in the reconstructed images which is what we want. This means our reconstruction loss will be higher for the CC images, therefore flagging them as possible anomalous data.

To summarise, the evaluation metric and band combination that maximises the difference in loss between the CC and non-CC test samples is the MSE $u-$model loss though there is not a significant difference between the MSE and L1 u-$model$ performances. Additionally, the samples cannot easily be differentiated based on their losses alone which means we must take a more sophisticated approach for classification which is why we utilise the K-means classifier. Specifically, we test the MSE loss values from the $u-$model reconstructed images (herein, $u-$loss) and similarly we confirm that MSE performs similarly to the L1 $u-$band loss values for the classification of our CC and non-CC samples and any choice between the two losses is arbitrary. This is shown in Table.~\ref{tab:u-band_loss_comp}, in which the MSE model has higher, accuracy and precision whereas the L1 model has higher recall and an F1-score.

\begin{table}
    \caption{Accuracy, precision, recall and F1-score evaluation metrics to compare the performance of K-means clustering algorithm when using different $u-$model losses combined with $u-g$ and $g-r$ colours.}\label{tab:u-band_loss_comp}
    \begin{tabular}{c|cccc}
        $u-$band Model Loss & Accuracy & Precision & Recall & F1-Score \\
        \hline
        MSE & 0.819 & 0.816 & 0.824 & 0.82 \\
        L1 & 0.817 & 0.80 & 0.844 & 0.822 \\
    \end{tabular}
\end{table}

\subsection{Comparison with Colour}
To maximise the degree of separation between the CC and non-CC samples, we compare different photometric colours provided by SDSS in tandem with the $u-$model MSE loss to try and find some defining relationship. Galaxy colours are already widely used to classify galaxies as star-forming or quiescent, LTG or ETG but are not sensitive enough to classify CCs. This is why they are used alongside other properties and spectral features such as [OII] and H$\alpha$ emission and H$\delta$ absorption when making selection cuts \citep{2018ApJ...868...84M}. By comparing the reconstruction loss (i.e. the outlier detection) with colour, we aim to add an extra layer of sensitivity that will diversify our threshold criteria. 

\begin{figure*}
    \centering
    \includegraphics[width=0.7\linewidth]{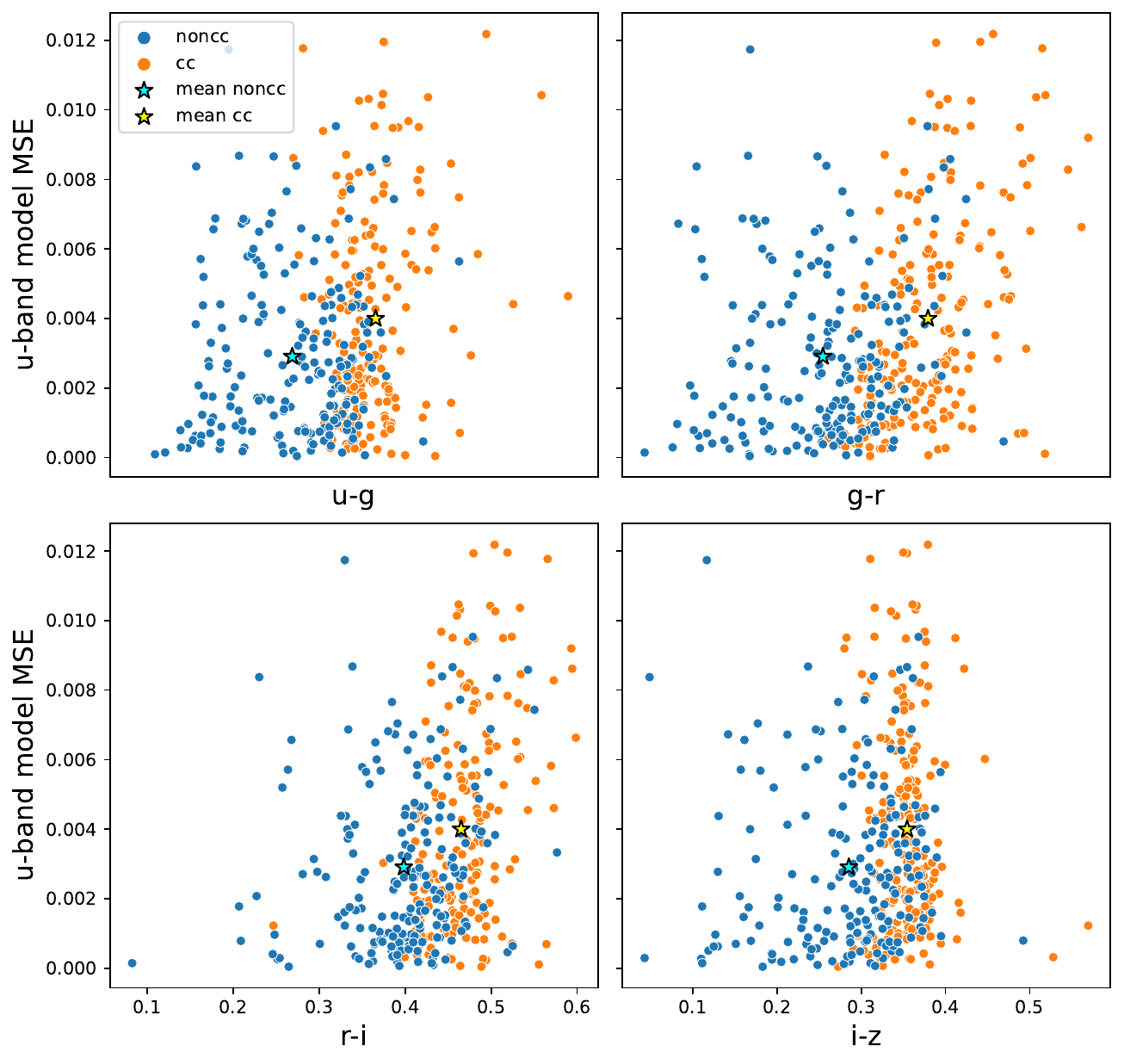}
    \caption{$u-g$, $g-r$, $r-i$ and $i-z$ photometric colours plotted against the $u-$model MSE loss for CC and non-CC galaxies. The CC sample is shown in orange and is typically redder than the blue non-CC sample. The most appreciable separation is apparent in the $u-g$ and $g-r$ colours. The blue and yellow stars correspond to the mean values for the non-CC and CC samples respectively.}
    \label{fig:2x2_col}
\end{figure*}

\begin{figure}
    \centering
    \includegraphics[width=\linewidth]{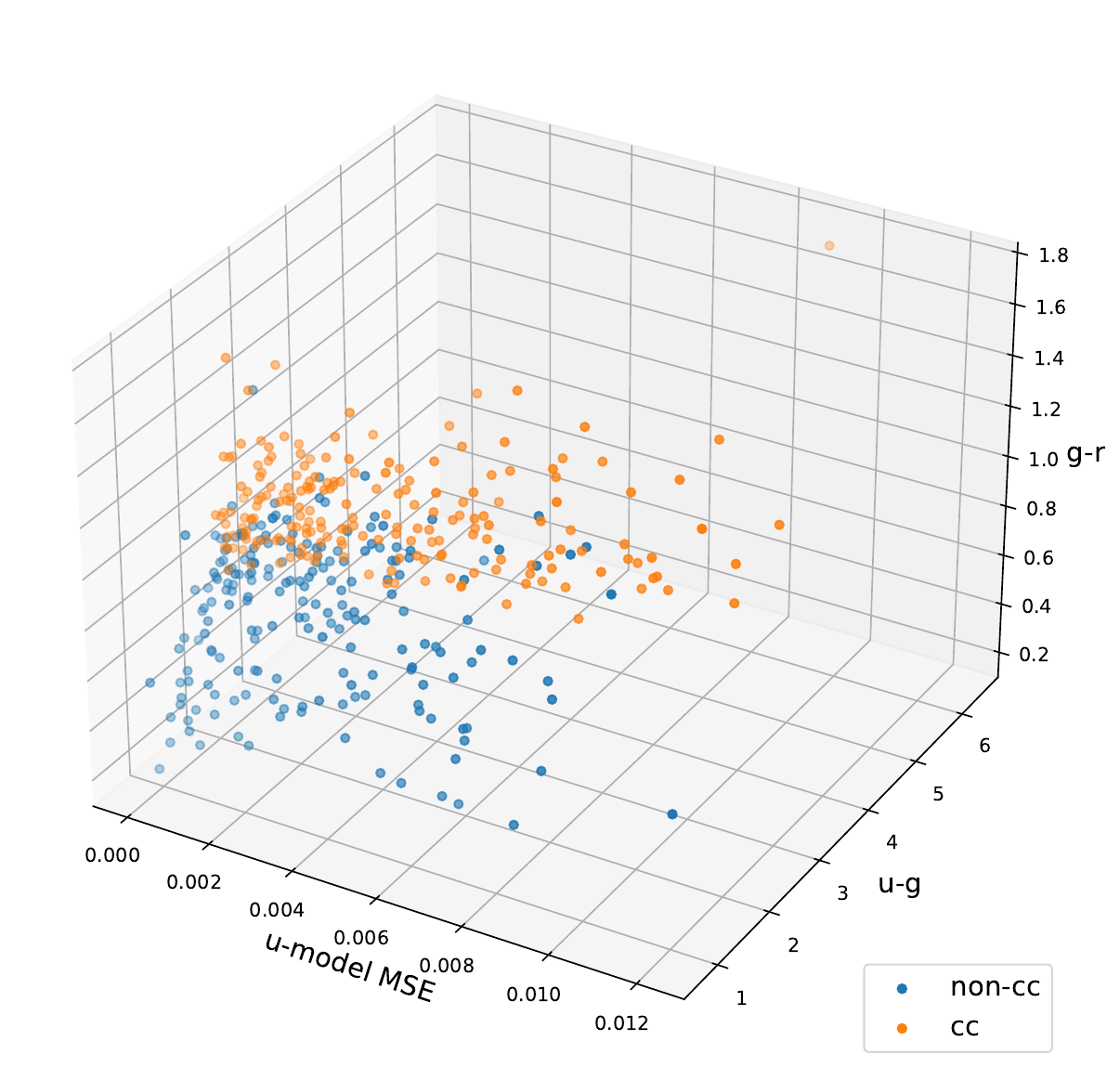}
    \caption{The 3D scatter plot shows a good amount of separation between the CC and non-CC samples when comparing the $u-$model MSE, $u-g$ and $g-r$ colours.}
    \label{fig:3d_mse_col}
\end{figure}

To begin, we compare the $u-g$, $g-r$, $r-i$ and $i-z$ colours against the $u-$model MSE losses in Fig.~\ref{fig:2x2_col} which shows there is a noticeable difference between the CC and non-CC samples. The CC galaxies all appear to be redder than the non-CC sample which is in line with the expected outcome according to the literature. The $u-g$ and $g-r$ colours show the largest amount of separation with both the sample distributions and mean values. Fig.~\ref{fig:3d_mse_col} shows the comparison of the $u-$model MSE, $u-g$ and $g-r$ colours which demonstrates a good amount of separation though there is still some amount of mixing at the boundary between the CC and non-CC samples. To separate the intermediate points something more complex than a simple threshold value is needed, such as an ML based clustering algorithm such as a K-means classifier.

\section{Classification Results} \label{class_results}
To classify the CC and non-CC samples using multiple metrics we use the K-means clustering classification to find the more complex relationships between multi-dimensional data. We aim to determine whether the inclusion of CAE losses improves the classification performance as opposed to using colour alone. To achieve this, we train 24 K-means models, half of which include the $u-$model MSE reconstruction losses and half are based on colour alone in order to make direct comparisons. We show an illustrative combination of colours that balances the importance between bluer and redder features, though this is not exhaustive of all combinations. 

\begin{figure}
    \centering
    \includegraphics[width=\linewidth]{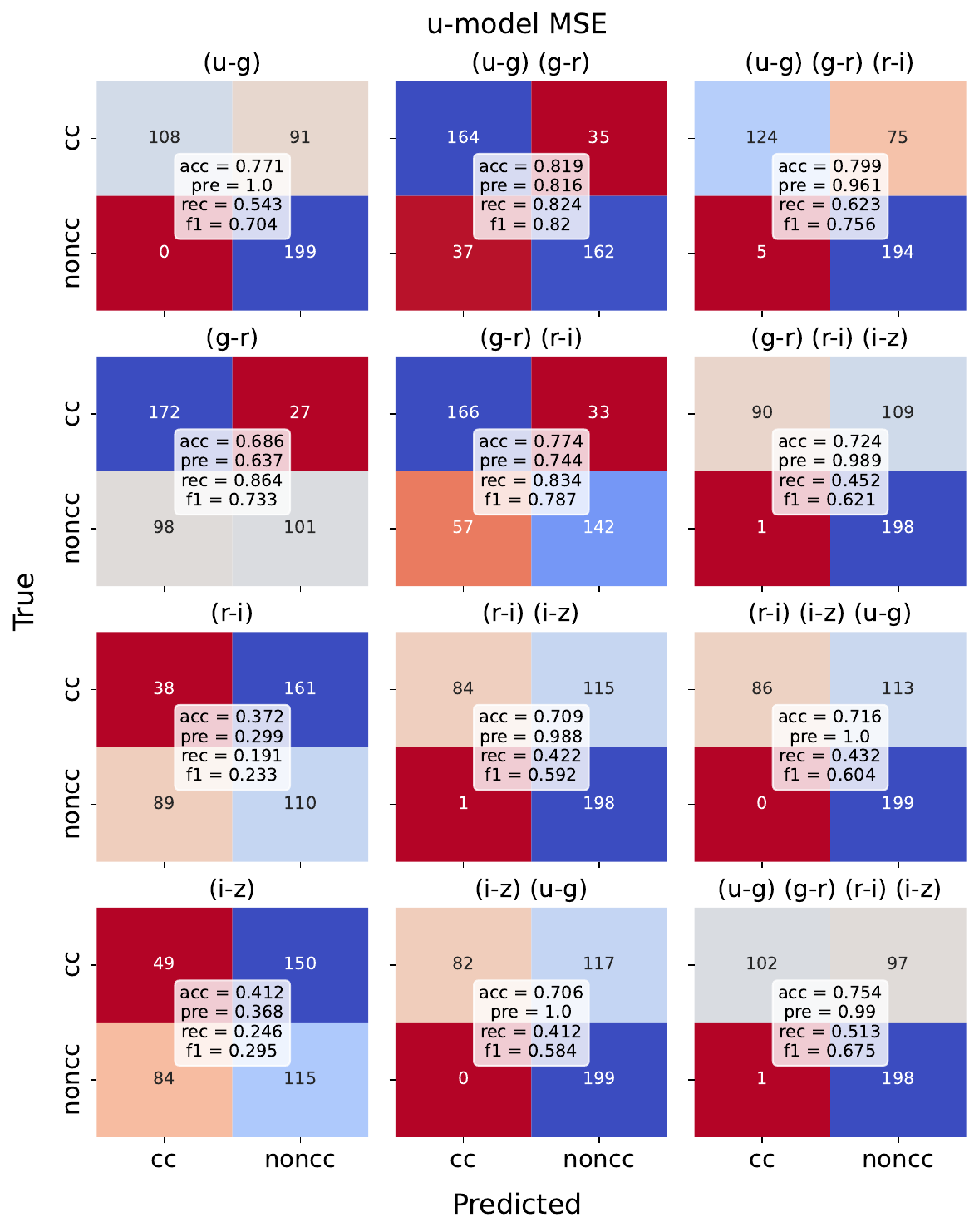}
    \caption[Confusion matrices of K-means predictions]{The confusion matrices of all 12 K-means models trained with a combination of different photometric colours and the $u-$model MSE. The best model is the $u-g$+$g-r$+MSE model as it has the most even predictions. The $g-r$+$r-i$+MSE and $g-r$+MSE also perform well, with only 33 and 27 missed CCs respectively. However, they both feature a high number of non-CC galaxies misclassified as CCs.}
    \label{fig:big_col_mse}
\end{figure}

\begin{figure}
    \centering
    \includegraphics[width=\linewidth]{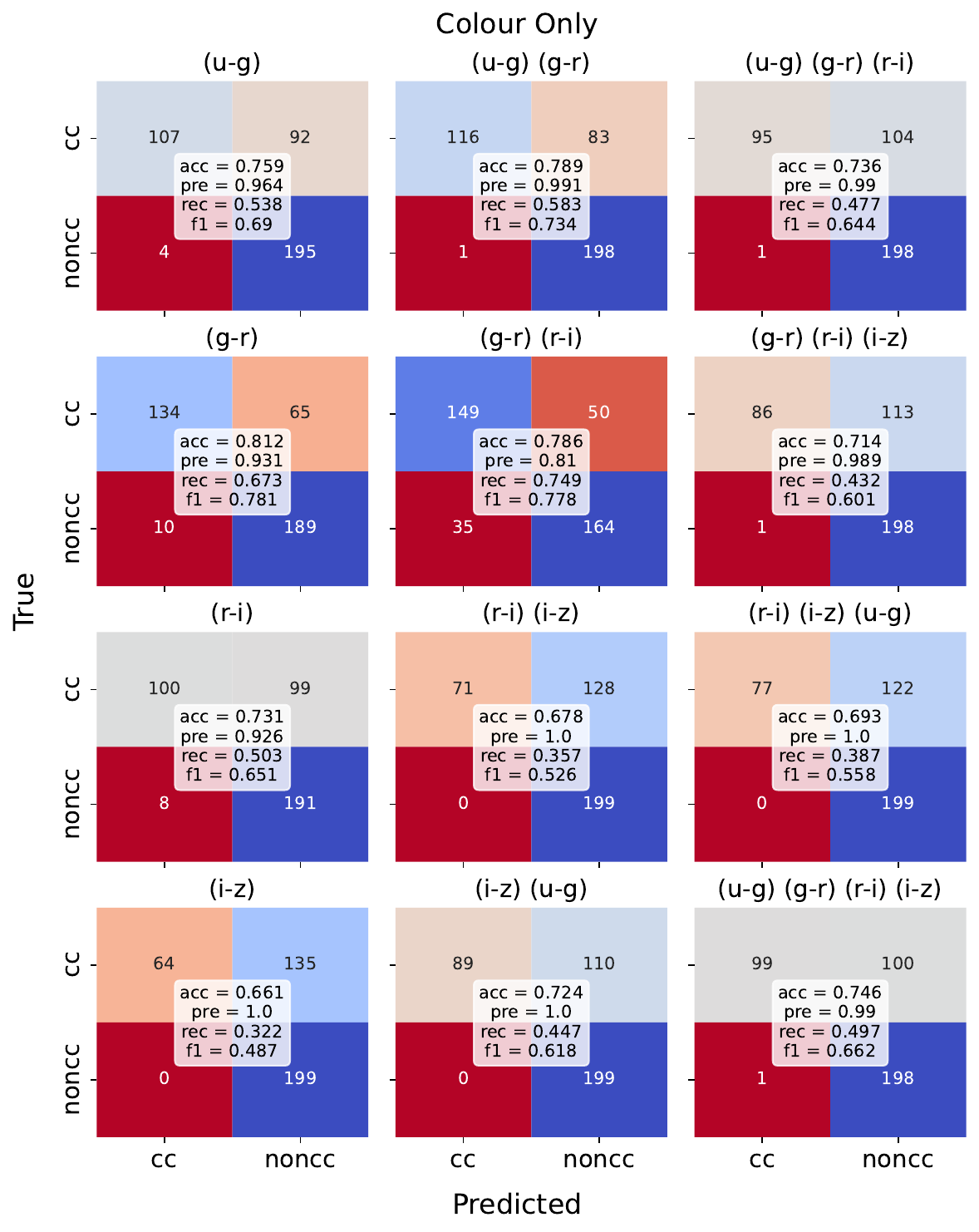}
    \caption[Colour-only confusion matrices]{The colour-only results are shown plotted with confusion matrices. Overall, these models perform worse than the colour-MSE models. Almost all of the colour-only models correctly classify all of the non-CCs but do so by missing a significant fraction of the CCs sample.}
    \label{fig:big_no_mse}
\end{figure}

We show the confusion matrices of the classification results for the colour+MSE models in Fig.~\ref{fig:big_col_mse} and the colour only models in Fig.~\ref{fig:big_no_mse} alongside the accuracy, precision, recall and F1-scores. In the context of our work, the most important metrics to score highly in are accuracy and recall as they focus on the overall correct predictions and false negative rate. This is because the overall aim is to use photometry and galaxy images in order to flag potential CCs as spectroscopic targets for further analysis in future spectroscopic surveys. Therefore, recall is especially important because we do not want to misclassify CCs as non-CCs whereas precision is less important because additional spectroscopic analysis should be performed to rule out any misclassified non-CCs. 

Fig.~\ref{fig:big_col_mse} shows that the ($u-g$)+($g-r$)+MSE K-means model performs symmetrically, with only 35 CCs being misclassified as non-CC and 37 non-CCs being misclassified as CC. Two of the worst models are the $r-i$ and MSE model and the $i-z$ and MSE model as they miss a large fraction of CCs whilst flagging a lot of non-CCs as CC candidates. Fig.~\ref{fig:big_no_mse} shows that the colour-only models perform significantly worse when misclassifying CCs as non-CCs. For example, the ($u-g$)+($g-r$) model misses 83 CCs in comparison to the 35 that the MSE version misses. Overall, when the bluer $u-g$ and $g-r$ colours are included in the input features the performance is better than when the redder $r-i$ and $i-z$ colours are included. 

\begin{figure}
    \centering
    \includegraphics[width=\linewidth]{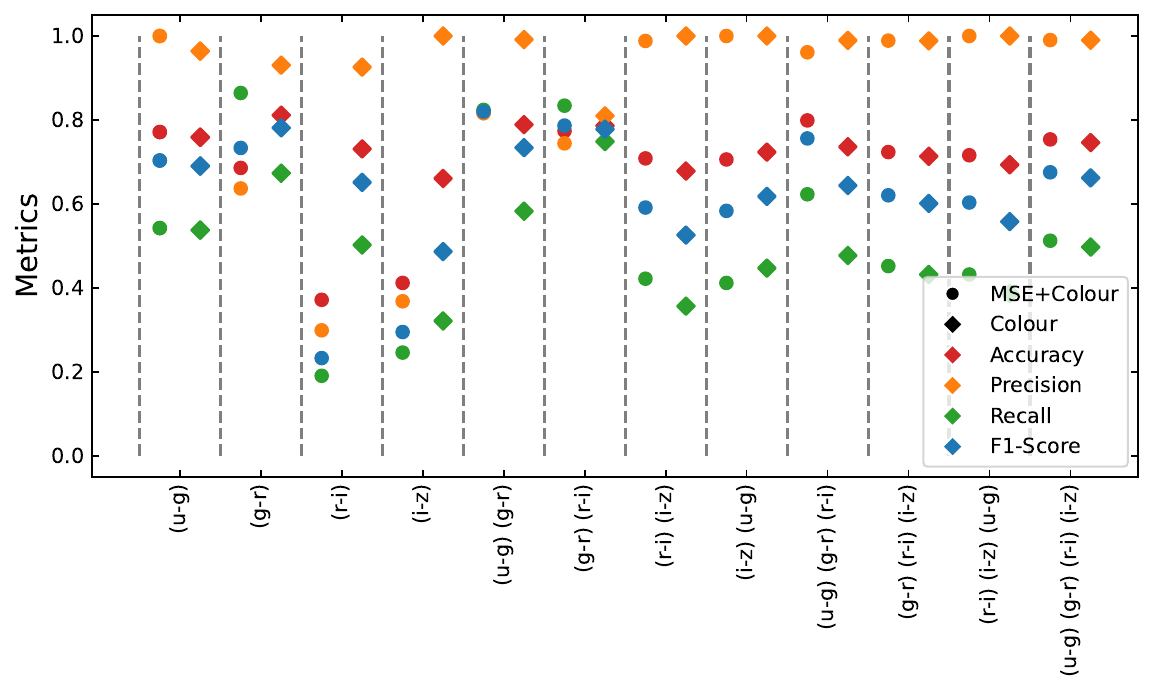}
    \caption{Each colour-only model (filled diamonds) is compared directly against its colour+MSE (filled circles) counterpart, separated by the dashed lines. The accuracy, precision, recall and F1-scores are shown in red, orange, green and blue respectively. The closer to one each metric is, the better the score is. }
    \label{fig:direct_comp}
\end{figure}

Additionally, we directly compare the classification metrics in Fig.~\ref{fig:direct_comp} across the colour+MSE and colour-only models. Almost all of the models feature high precision metrics because the rate of false positives is low, this means almost all non-CCs are correctly classified. However, this appears to be because the models are classifying everything as non-CC, including our CC samples which is evident when we compare the recall which quantifies the rate of true positives. A low recall value means the model is misclassifying CCs as non-CCs which is the worst possible outcome for the purpose of our work. There is only one colour-only model and three colour+MSE models that produce relatively high recall values; the $(g-r)$+$(r-i)$, $(g-r)$+MSE, $(u-g)$+$(g-r)$+MSE and $(g-r)$+$(r-i)$+MSE models. The $(u-g)$+$(g-r)$+MSE model performs especially well because all calculated metrics are approximately the same, showing the model is the most well-rounded. 

\begin{figure}
    \centering
    \includegraphics[width=\linewidth]{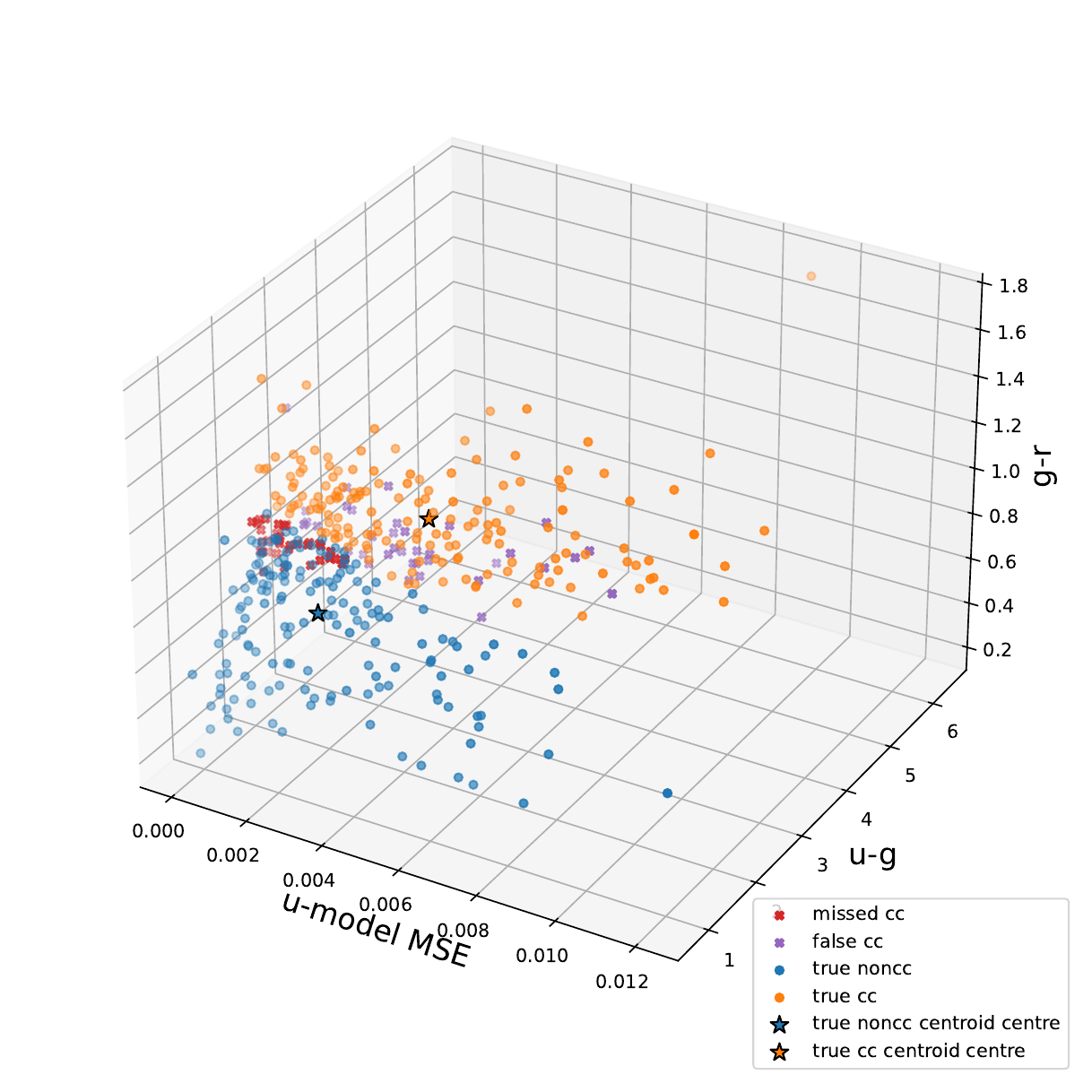}
    \caption{The CC and non-CC cluster centres are shown with the red and blue stars respectively. Correctly classified CCs and non-CCs are shown in orange and blue circles. The non-CC galaxies that are misclassified as CCs are shown with purple whereas the CCs misclassified as non-CCs are shown as red crosses.}
    \label{fig:kmeans_3d}
\end{figure}

\begin{figure}
    \centering
    \includegraphics[width=\linewidth]{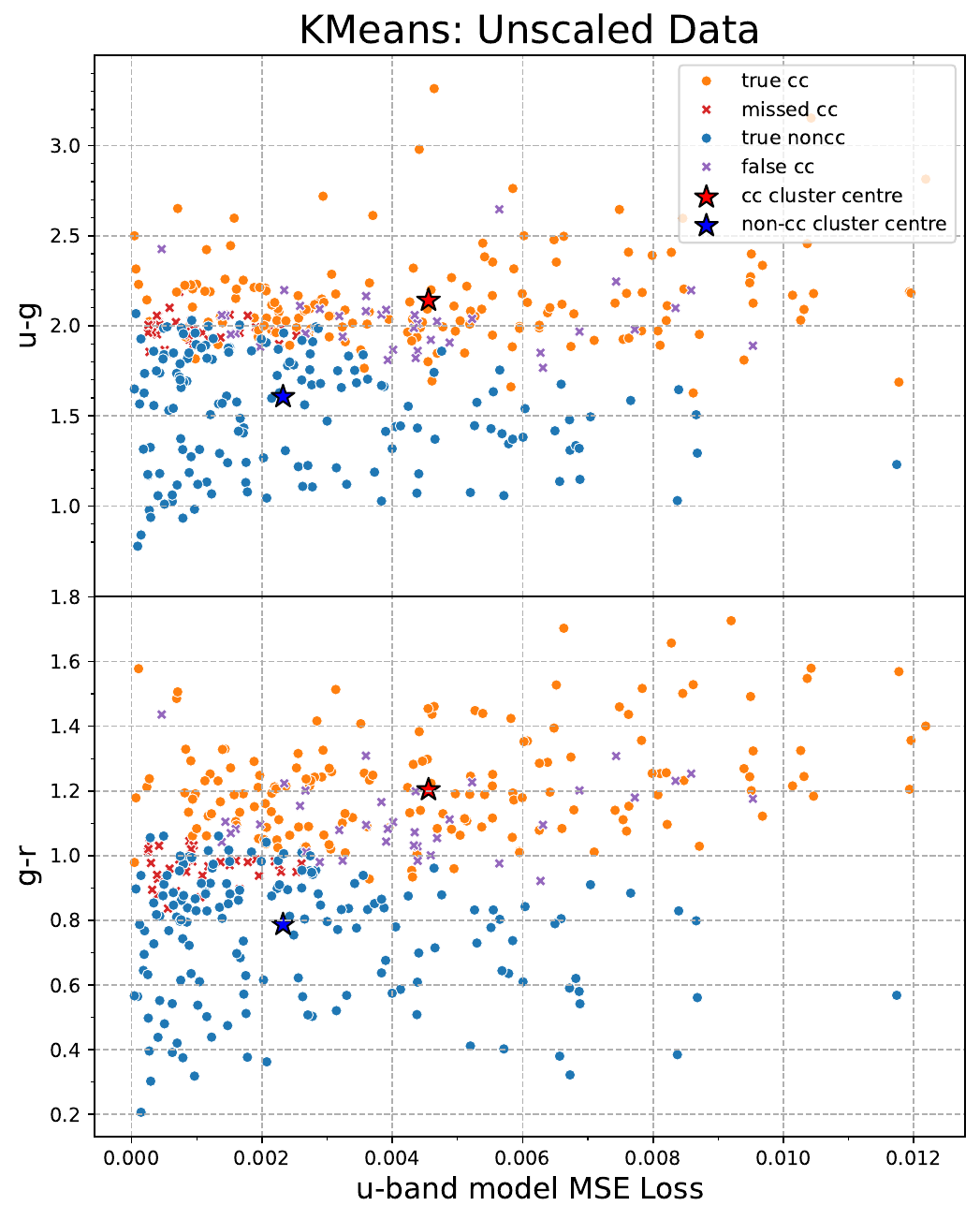}
    \caption{The same marker convention as Fig.~\ref{fig:kmeans_3d} shows the difference in the $u-g$ and $g-r$ colours in comparison with the $u-$model MSE values. For clarity, we unscale the data to show the true colour magnitudes.}
    \label{fig:kmeans_unscale}
\end{figure}

We show the $(u-g)$+$(g-r)$+MSE model classifications in Fig.~\ref{fig:kmeans_3d} to illustrate a broad view of CC and non-CC clusters. It is clear that the model struggles to differentiate galaxies along the boundary, as this is where the majority of missed CCs and false CCs are situated. Fig.~\ref{fig:kmeans_3d} shows that all of the missed CCs are in this boundary cluster. Additionally, Fig.~\ref{fig:kmeans_unscale} shows there are a number of non-CC galaxies incorrectly classified as CCs that permeate the true CC sample. This is most likely due to the nature of the K-means algorithm as it classifies points based on the distance to the nearest centroids. This means it performs best with data that does not overlap which is evident in our results. Other ML algorithms such as support vector machines perform better with overlapping data as they are able to find non-linear decision boundaries. However, they are supervised which means the labelled input data needs to be separated into training and validation sets. Additionally, supervised models required more data overall in order to train effectively which means our dataset of 398 galaxies is too small to evaluate a supervised model soundly.

\begin{figure}
    \centering
    \includegraphics[width=\linewidth]{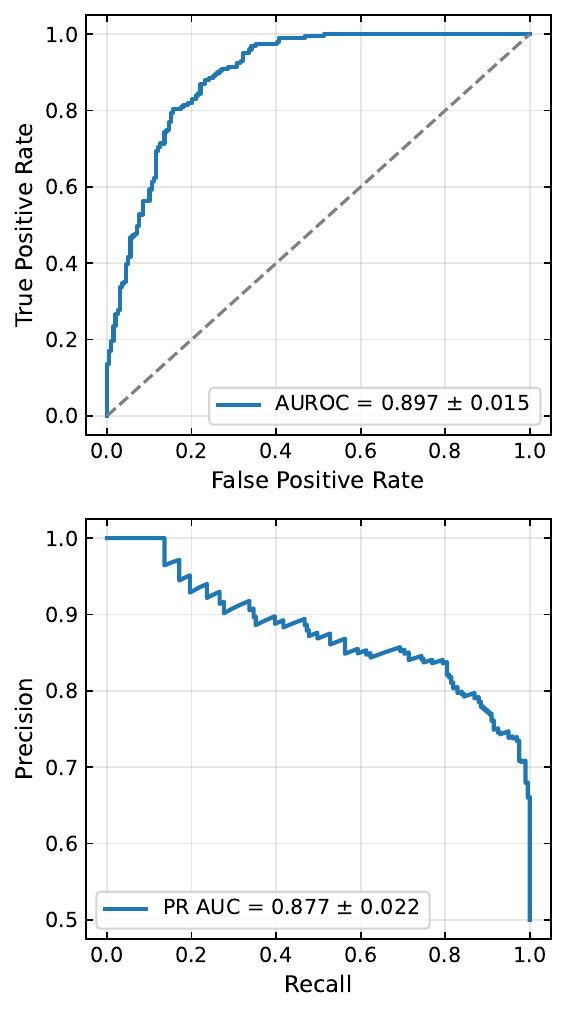}
    \caption{In the \textit{top panel} we show the ROC curve calculated using repeated stratified K-fold cross-validation for 5 folds and 10 repeats. Similarly, we show the precision-recall curve in the \textit{bottom panel} for the same 5 folds and 10 repeats. For each curve we also show the AUC.}
    \label{fig:roc_pr_curves}
\end{figure}

To confirm the K-means model results are not optimistically biased because of the input sample we perform repeated stratified K-fold cross-validation to randomly select different splits and evaluate how this affects the model. We perform 10 repeats using 5 folds each and plot the resulting ROC and precision-recall curves in Fig.~\ref{fig:roc_pr_curves} alongside the AUROC and PR-AUC respectively. We find the AUROC~$=0.897\pm0.015 (95\%$~CI~$ 0.865-0.926)$ and PR-AUC~$=0.877\pm0.022(95\%$~CI~$0.834-0.918)$ which shows the $(u-g)$+$(g-r)$+MSE model is stable with respect to different data splits and we can confidently say the results are not optimistically biased by a favourable data split.

\subsection{Analysis of Misclassifications}
\begin{figure}
    \centering
    \includegraphics[width=\linewidth]{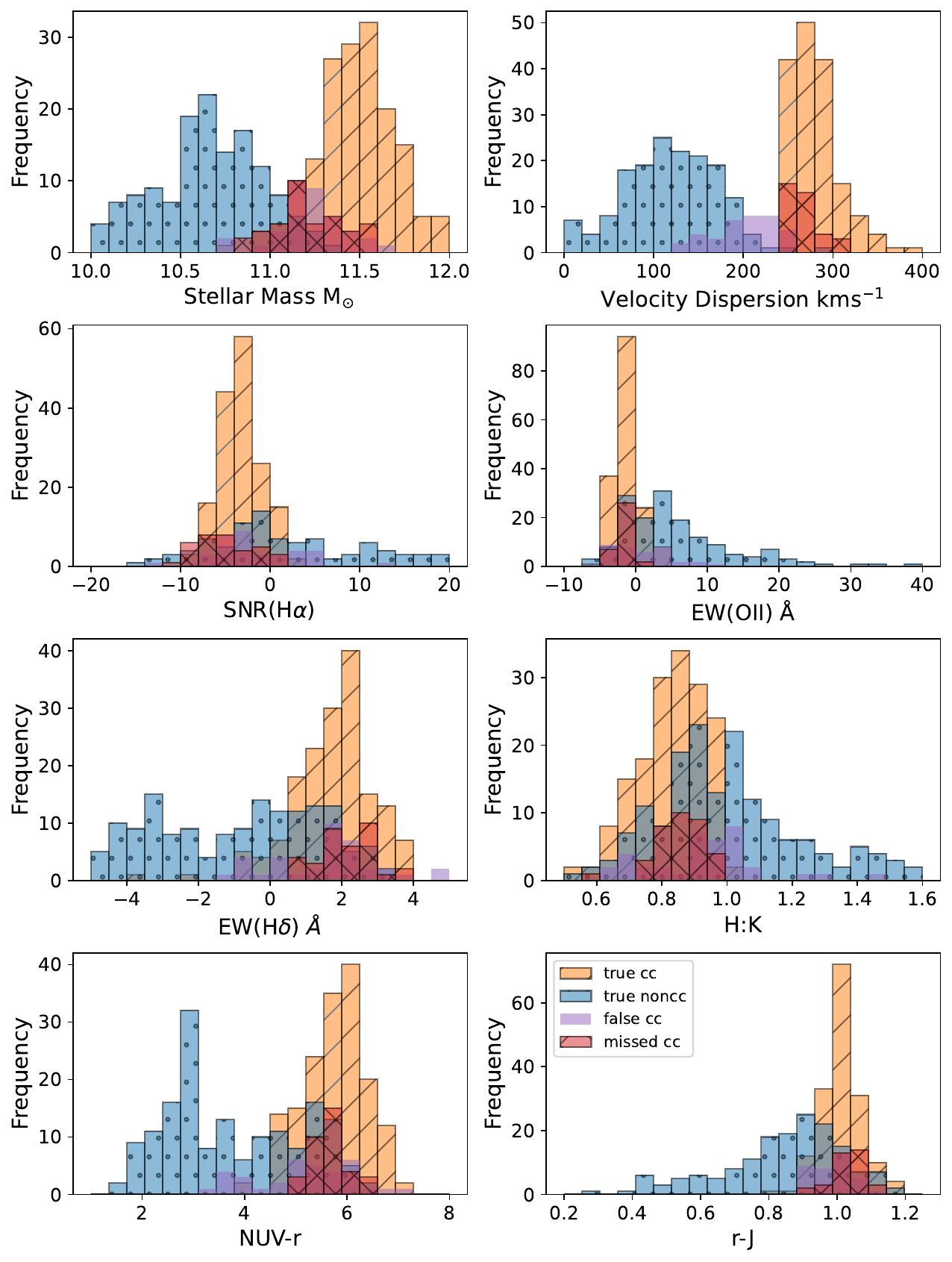}
    \caption{Histograms show the distribution of different classifications across the different selection cuts. True CCs are shown in orange and true non-CCs are shown in blue. Non-CCs falsely classified as CCs are shown in purple whereas CCs classed as non-CCs are shown in red. }
    \label{fig:class_anal}
\end{figure}

Fig.~\ref{fig:class_anal} shows the spread of classifications with respect to the selection cuts used to create the CC sample described in section~\ref{data}. The non-CC sample shares some overlap with the CC sample for a number of properties, such as the SNR(H$\alpha$), EW([OII]) and EW(H$\delta$) spectral lines, the H:K ratio and the (NUV$-rJ$) and  ($r-J$) colours, this just demonstrates that the non-CCs in the ``overlap'' have failed at least one of the selection cuts and therefore been discarded as CC candidates. In addition, all misclassified galaxies fall in these overlapping ranges which suggests these galaxies exhibit intermediate spectral features and galaxy properties making them harder for the model to distinguish. The only exception to this is the stellar velocity dispersion cut which shows a clear separation between the CC and non-CC samples with a small bridging population of false non-CCs.

\begin{figure}
    \centering
    \includegraphics[width=\linewidth]{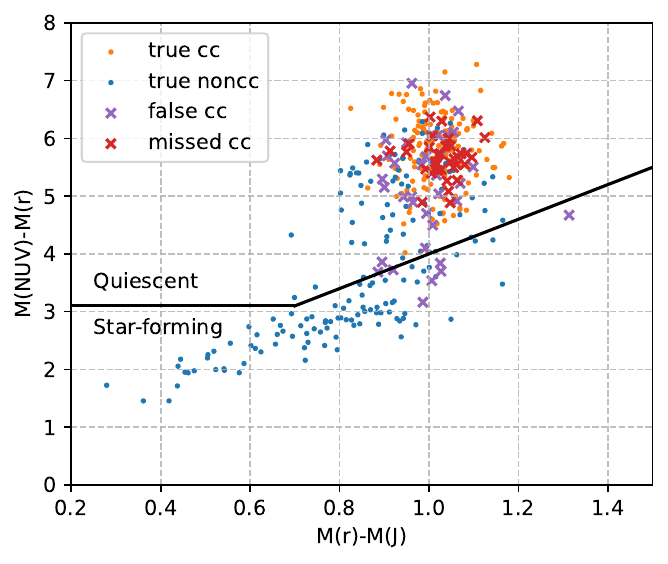}
    \caption{The NUV$-r$ and $r-J$ colours are plotted to compare the number of non-CCs and CCs across the star-forming and quiescent regions of the figure. True non-CCs and CCs are shown with blue and orange points, whereas missed CCs and false CCs are shown with red and purple crosses.}
    \label{fig:class_nuvrj}
\end{figure}

In addition to comparing the values needed for making the selection cuts shown in Fig.~\ref{fig:class_anal}, we plot the NUV$rJ$ colour magnitude diagram in Fig.~\ref{fig:class_nuvrj}. Galaxies above $\text{NUV} - r>3(\text{NUV} - J)+1$ and $\text{NUV} - r>3.1$ are considered quiescent whereas galaxies below this line are typically star-forming \citep{2013A&A...556A..55I}. All of our galaxies that fall below the line are non-CCs, mostly true non-CCs but there are a few false CCs close to the boundary though there are a number of non-CCs and false CCs above the line. All missed CCs are concentrated on the main CC clump indicating that they are quiescent. This is reflected in Fig.~\ref{fig:class_anal} as the missed CC sample heavily overlaps the true CCs. It is important to note that just because a galaxy is above the line does not mean it should be a CC as it must pass all selection cuts in addition to the NUV$rJ$ cut. 

\begin{figure}
    \centering
    \includegraphics[width=\linewidth]{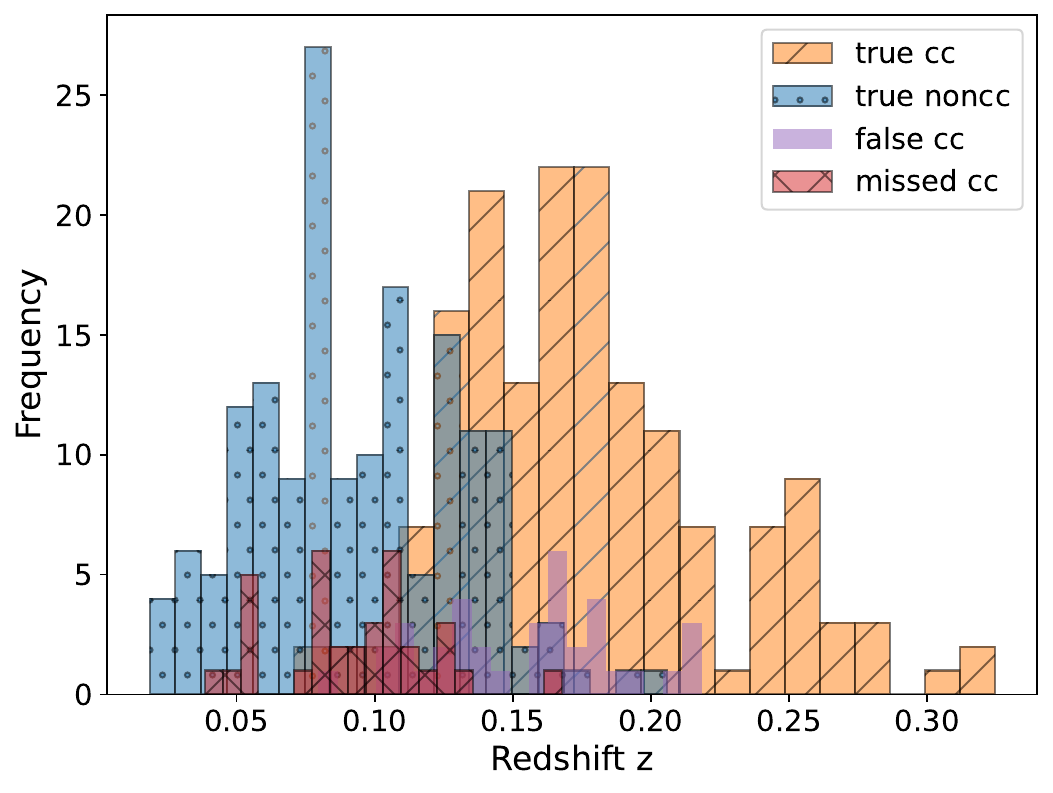}
    \caption[Histogram of the redshifts of subsamples]{The redshift range for each correctly classified CC (in orange), non-CC (in blue) and misclassified false CC (in purple) and missed CC (in red). The CC sample has a higher redshift range than the non-CC sample and the missed CCs fall on the low end of the true CC range.}
    \label{fig:class_z}
\end{figure}

Fig.~\ref{fig:class_z} compares the redshift range of the CC and non-CC populations and shows the missed CCs all fall on the low end of the CC range ($0.03\lesssim z \lesssim 0.16$). When performing the selection cuts and creating the non-CC test sample we do not make any selections based on redshift. 

\begin{figure}
    \centering
    \includegraphics[width=\linewidth]{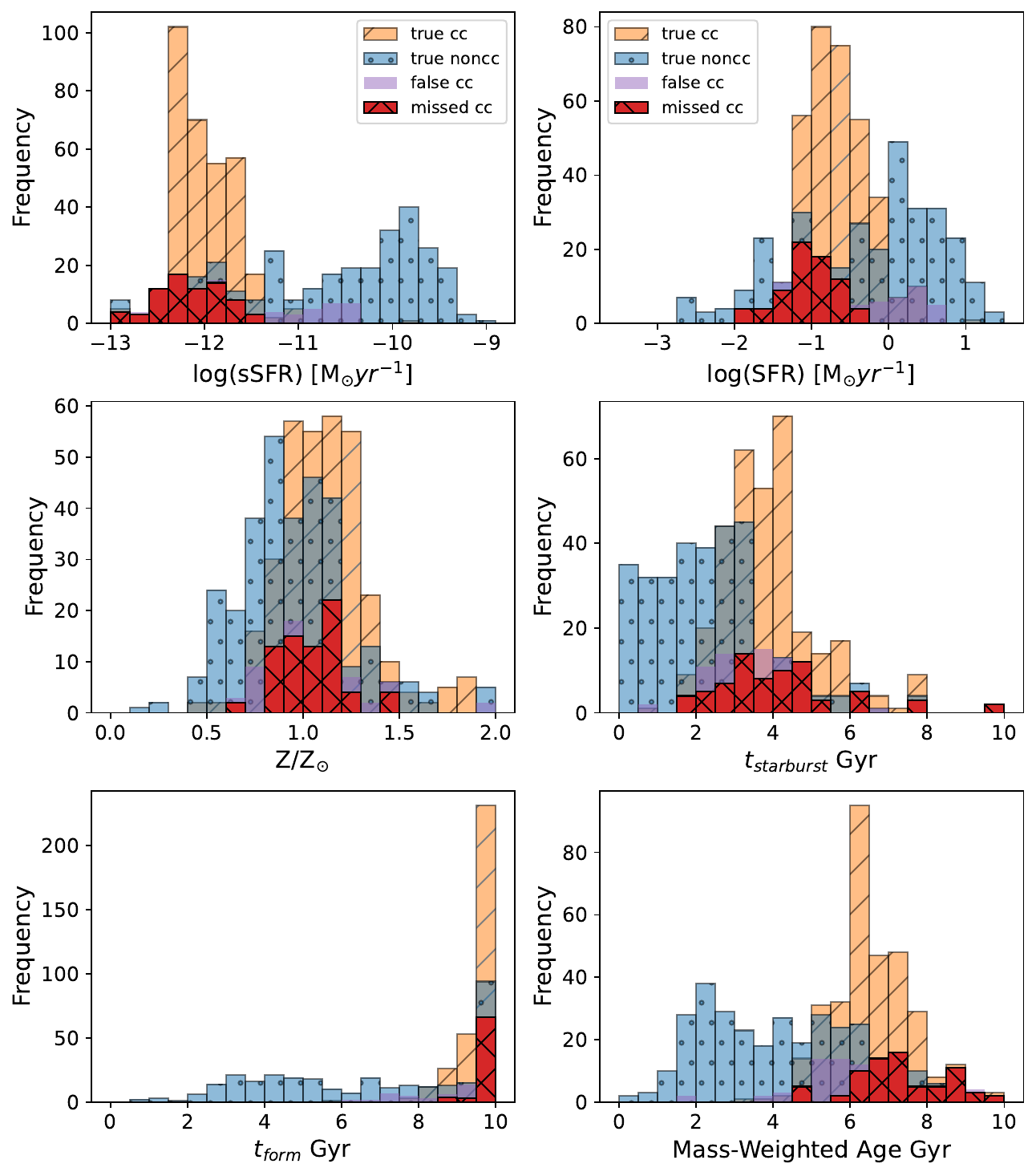}
    \caption[Histograms of subsamples and their properties]{The top two panels show the distribution of classes with respect to the sSFR (\textit{left}) and SFR (\textit{right}) for each galaxy. The middle two panels show the metallicity (\textit{left}) and the time since the most recent starburst (\textit{right}). Finally, the bottom two panels show the ages of the oldest stars in the galaxies (\textit{left}) and the median-mass weighted age (\textit{right}).}
    \label{fig:class_props}
\end{figure}

Finally, we compare the different predicted classes with galaxy properties predicted by the MagPhys spectral fitting code \citep{2008MNRAS.388.1595D} in Fig.~\ref{fig:class_props}. All properties derived from MagPhys are given as a median most likely value for each galaxy. Similarly to our previous investigations of the misclassified CCs, the missed CCs all reside in the overlap between the true CC and non-CC sample. The true CC sample are typically the oldest galaxies with the lowest sSFRs and the oldest stellar populations t$_{\text{form}}$. 

In order to understand why the K-means clustering algorithm misclassifies some CCs and non-CCs we show the thumbnail images of all missed CCs, false CCs and a sample of true CC and non-CCs in order to evaluate what similarities or differences may be present. As the K-means algorithm uses the $u-$model CAE losses we show the $u-$band images in addition to the $r-$band images as this band shows the morphology more clearly. 

\begin{figure}
    \centering
    \includegraphics[width=\linewidth]{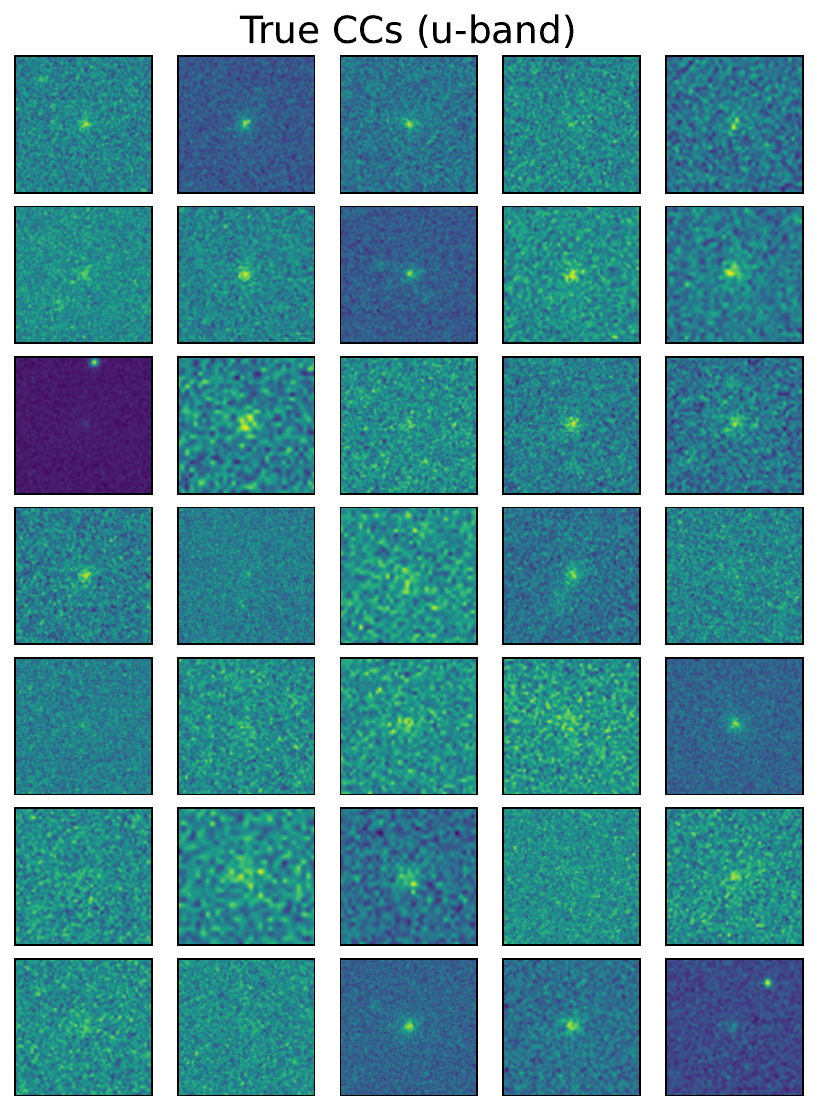}
    \caption[The $u-$band photometric images of 35 random correctly classified true CCs]{The $u-$band photometric images of 35 random correctly classified true CCs to show what we expect to see when observing CCs. Typically, the CCs are very faint or invisible when compared to the background noise. For the CCs that do have visible UV emission, it is mostly concentrated to the galactic centre as opposed to Fig.~\ref{fig:truenon_uband} which shows spiral galaxies with the UV emission in their arms that is indicative of star-formation.}
    \label{fig:true_uband}
\end{figure}

\begin{figure}
    \centering
    \includegraphics[width=\linewidth]{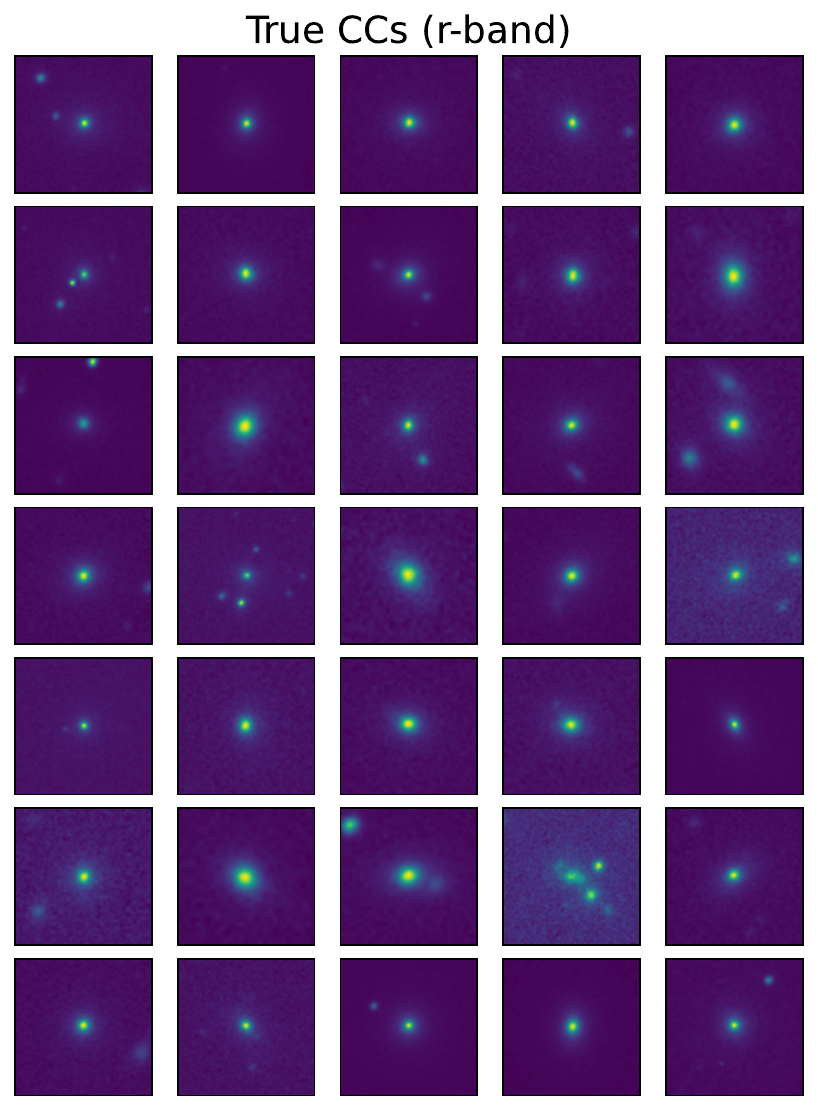}
    \caption[The $r-$band photometric images of 35 random correctly classified true CCs]{The $r-$band photometric images of 35 random correctly classified true CCs to show what we expect to see when observing CCs. All CCs appear to be E0-E5 in morphology with a large number of them having neighbours.}
    \label{fig:true_rband}
\end{figure}

The illustrative sample of correctly classified CCs in the $u-$band in Fig.~\ref{fig:true_uband} and $r-$band in  Fig.~\ref{fig:true_rband} shows mostly elliptical E0-E5 target galaxies, commonly with neighbours which supports the idea that the most massive galaxies tend to be found in high-density regions. It is interesting to see that most of the neighbours are only visible in the $r-$band images and appear to be almost invisible in the $u-$band. In general, we expect the CCs to be very faint if not invisible in the $u-$band as they should be completely absent of UV emission caused by star-formation which is confirmed when inspecting Fig.~\ref{fig:true_uband}.


\begin{figure}
    \centering
    \includegraphics[width=\linewidth]{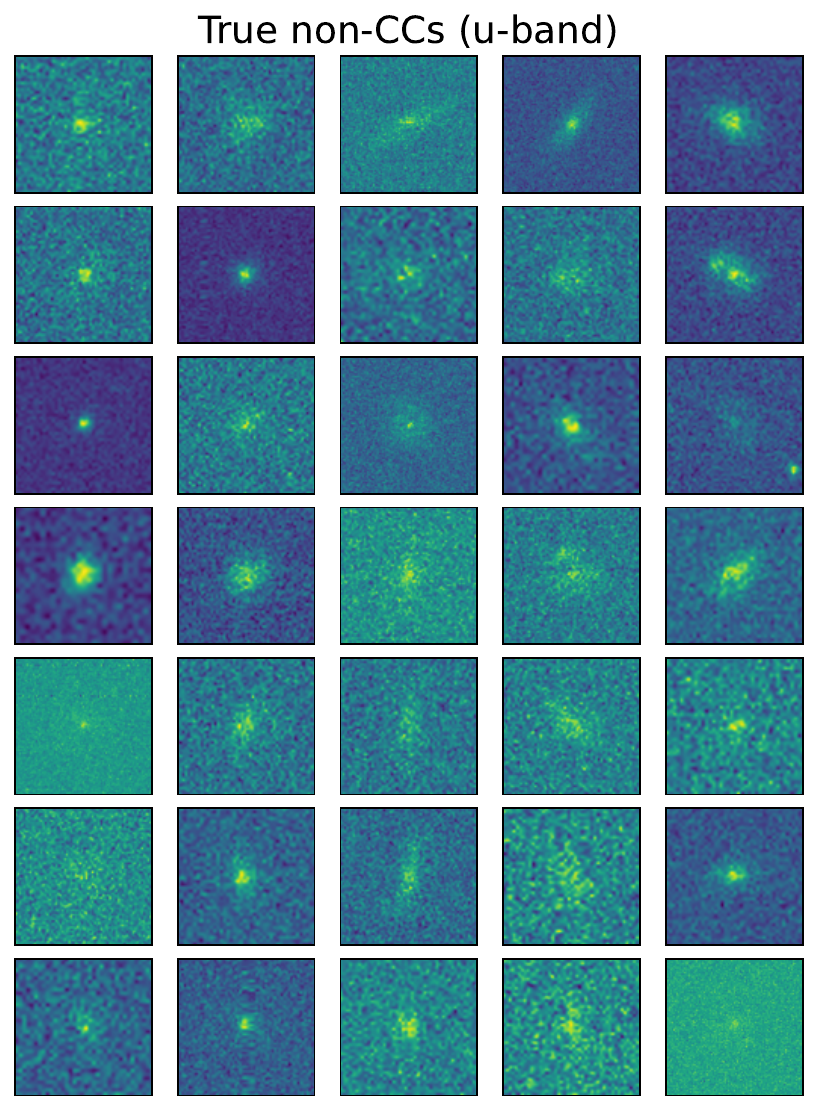}
    \caption[The $u-$band photometric images of 35 random correctly classified true non-CCs]{The $u-$band photometric images of 35 random correctly classified true non-CCs to show what we expect to see when observing non-CCs. Some of the non-CCs hint at their morphology in the $u-$band as UV emission can be seen where the spiral arms are present in Fig.~\ref{fig:truenon_rband}.}
    \label{fig:truenon_uband}
\end{figure}

\begin{figure}
    \centering
    \includegraphics[width=\linewidth]{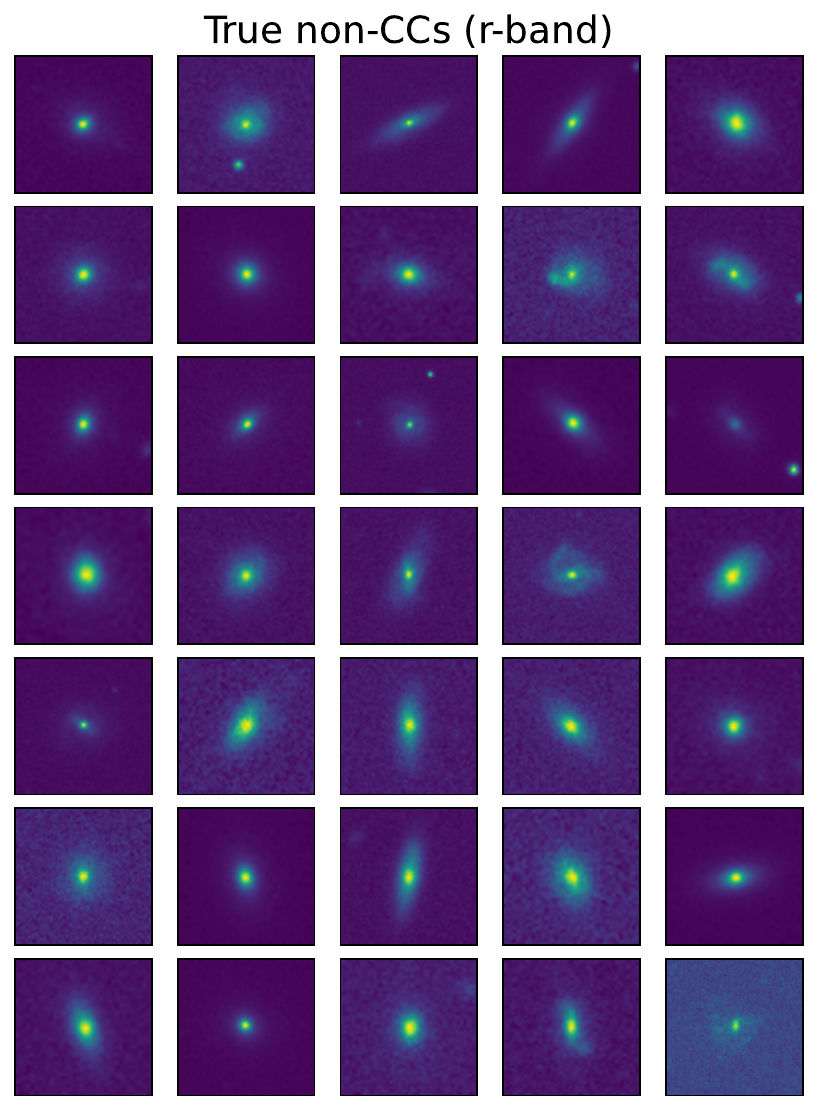}
    \caption[The $r-$band photometric images of 35 random correctly classified true non-CCs]{The $r-$band photometric images of 35 random correctly classified true non-CCs to show what we expect to see when observing non-CCs.}
    \label{fig:truenon_rband}
\end{figure}

In comparison, Fig.~\ref{fig:truenon_uband} and Fig.~\ref{fig:truenon_rband} show the $u-$band and $r-$band images of the correctly classified non-CC sample. Though the non-CCs are also fainter in the $u-$band in comparison to the $r-$band, by eye, they are typically much brighter than the CC sample. In both Fig.~\ref{fig:truenon_uband} and Fig.~\ref{fig:truenon_rband} there are very few visible neighbours in comparison to the true CC sample. The majority of the non-CCs appear to be edge-on spirals, lenticular galaxies or face-on spirals with only a few examples that could be considered elliptical. 


\subsubsection{Visual Inspection of Misclassified Galaxies}
\begin{figure}
    \centering
    \includegraphics[width=\linewidth]{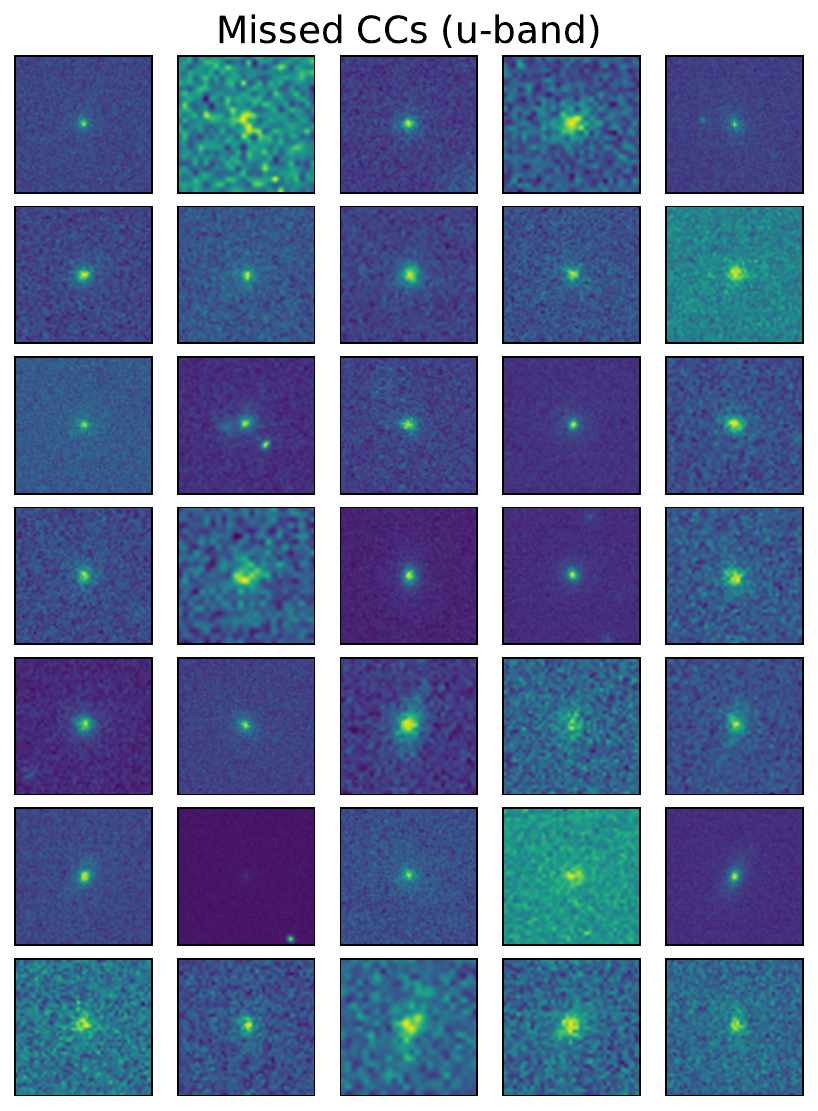}
    \caption{The $u-$band photometric images of all 35 misclassified true CCs.}
    \label{fig:missed_uband}
\end{figure}

\begin{figure}
    \centering
    \includegraphics[width=\linewidth]{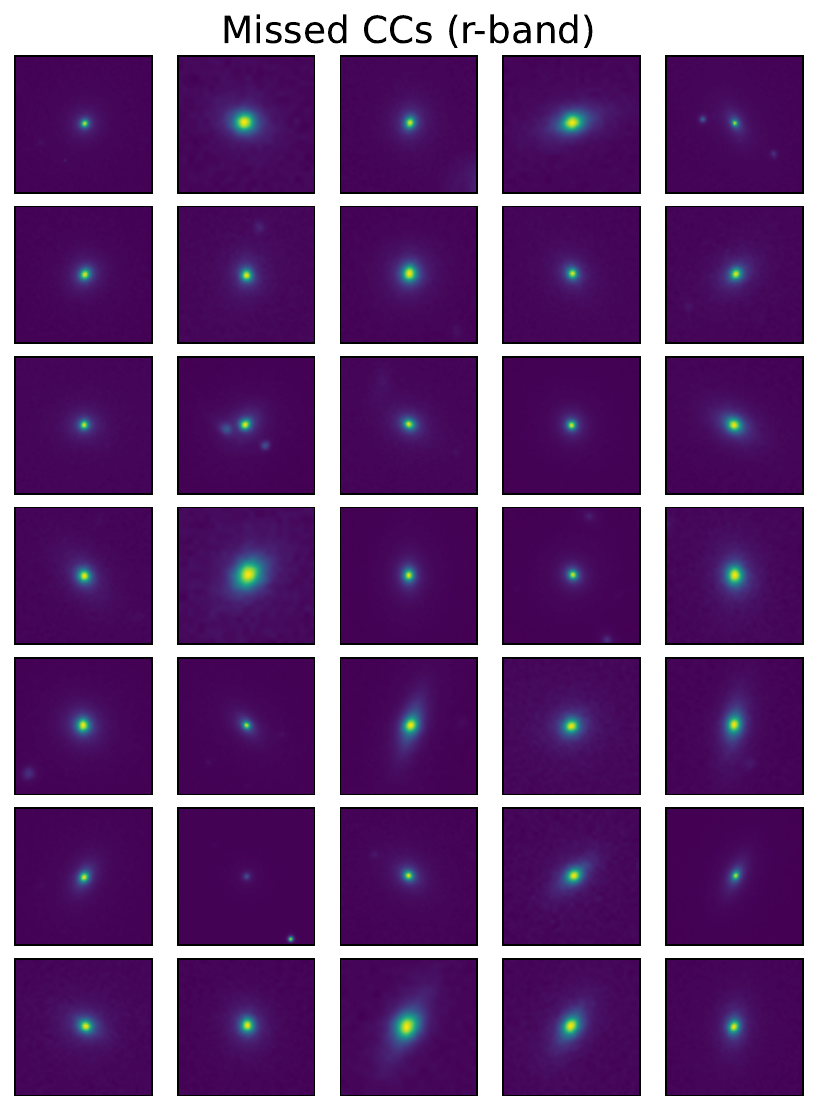}
    \caption{The $r-$band photometric images of all 35 misclassified true CCs.}
    \label{fig:missed_rband}
\end{figure}

Fig.~\ref{fig:missed_uband} and Fig.~\ref{fig:missed_rband} shows all 35 true CCs that the K-means algorithm misclassifies as non-CCs (missed CCs). Overall, by eye, the missed CCs seem to be brighter in the $u-$band in comparison with the correctly classified CCs as their background pixels show a larger contrast with the galaxy pixels. This makes them appear to be closer to the true non-CCs in the $u-$band. Additionally, there are fewer neighbours in the missed CC sample in comparison to the true CCs, this is apparent in both Fig.~\ref{fig:missed_uband} and Fig.~\ref{fig:missed_rband}. Finally, there are more examples of extended diffuse regions in Fig.~\ref{fig:missed_rband} which give some of the missed CCs a more extended elliptical/lenticular shape. Though this is more obvious in the $r-$band, it is also somewhat visible in the $u-$band for a few examples. 


\begin{figure}
    \centering
    \includegraphics[width=\linewidth]{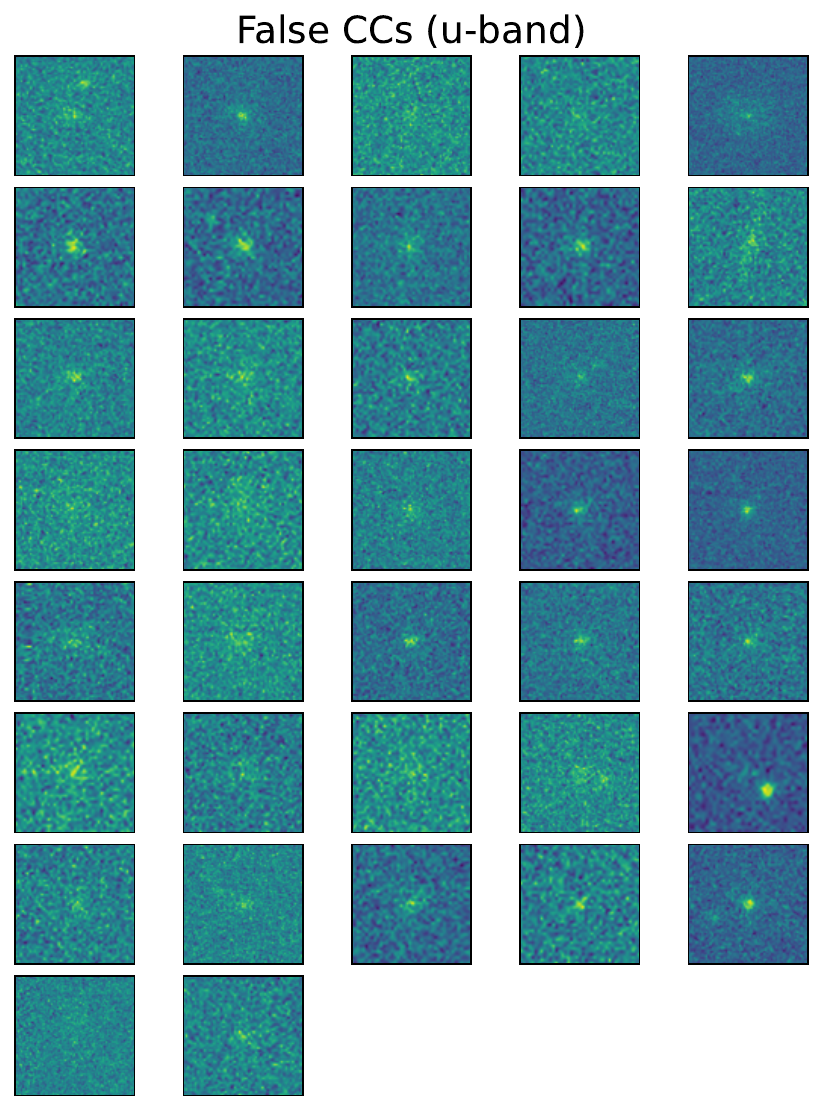}
    \caption{The $u-$band photometric images of all 37 misclassified true non-CCs.}
    \label{fig:false_uband}
\end{figure}

\begin{figure}
    \centering
    \includegraphics[width=\linewidth]{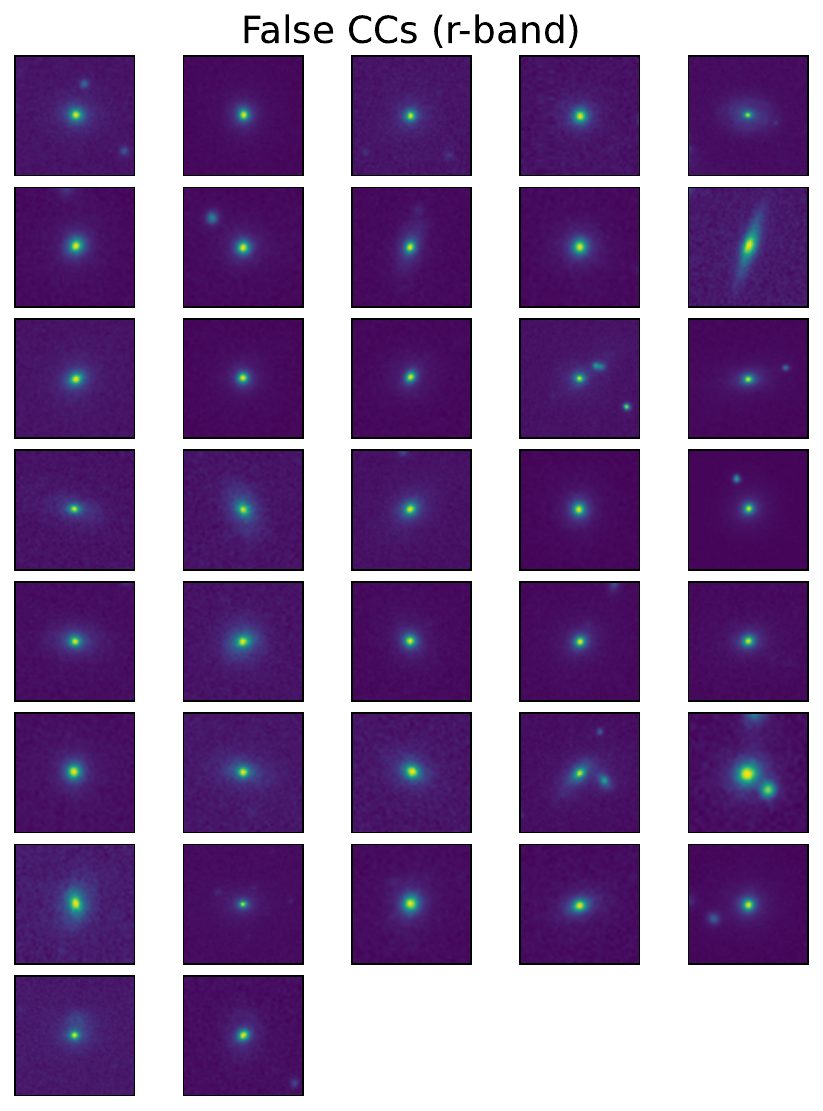}
    \caption{The $r-$band photometric images of all 37 misclassified true non-CCs.}
    \label{fig:false_rband}
\end{figure}

Finally, Fig.~\ref{fig:false_uband} and Fig.~\ref{fig:false_rband} shows all 37 true non-CCs that the K-means algorithm misclassifies as CCs (false CCs). The $u-$band images of the false CCs show a similar level of contrast to the true CCs in Fig.~\ref{fig:true_uband}, in which the target galaxies are very faint or even invisible when compared with the background. Additionally, in the $r-$band there are a number of false CCs that have neighbours and with the exception of one lenticular galaxy, all the false CCs appear to be elliptical in nature. 


By examining the misclassified galaxies in comparison with the correctly classified galaxies we provide context to the CAE $u-$band losses that are used for the K-means predictions. Because the missed CCs appear to be closer in appearance to the true non-CCs and the false CCs appear to be closer to the true CCs, this suggests the CAE is able to recreate the missed CC images better than the false CC images which leads to better loss scores for the false CCs. This means that when the K-means algorithm is fed the $u-$model MSE losses, it misclassifies the false CCs as CCs. This is compounded by the fact that the $u-g$ and $g-r$ colours that are also used in the K-means model fall in the overlapping boundary region for the misclassified galaxies.

Overall, our results show that the identification of cosmic chronometers using photometry alone is not an easy task, which is apparent in the current literature. However, our two-model framework is a useful tool to flag potential CC candidates at lower redshift for future spectroscopic surveys in which a more thorough analysis may be completed before using them to calculate the Hubble constant.

\section{Impact of CC Sample Contamination on Derived Measurements}
Here we compare the effect on the calculated value of the Hubble constant when using ML-predicted CCs versus traditionally selected CCs. The ML CC sample includes the 164 correctly classified CCs and the 72 misclassified true CCs and false CCs classified by the K-means algorithm. This will allow us to evaluate the classification performance of the model by comparing the predicted $H_0$ values and whether the ML model is practical for future studies. However, it is important to clarify that these measurements of $H_0$ are only for the purpose of comparing the subsamples and should not be taken as robust measurements because the uncertainty in age limits the number of redshift bins we can use. We are only able to use three redshift bins.

\begin{table}
    \centering
    \caption{We show the breakdown of each subsample used for comparing $H_0$ calculations. The number of galaxies in each sample is given for the whole sample in brackets with the number of cross-matched galaxies given next to it.}
    \begin{tabular}{c|cc}
        Subsample & Total No. \\
        \hline
        Correct (164) & 161 \\
        \hline
        Missed only (35) & 35 \\
        True CC + Missed (199) & 196 \\
        \hline
        False only (37) & 37 \\
        True CC + False (201) & 198 \\
        True non-CC + False (199) & 196 \\
        \hline
        All (398) & 392 \\
    \end{tabular}
    \label{tab:ccs_for_h0}
\end{table}

Table~\ref{tab:ccs_for_h0} shows the breakdown of the different subsamples which are cross-matched with the MagPhys table in the GAMA survey to ensure each galaxy in the sample has an age associated with it. The total number of galaxies in each sample is given in brackets in Table~\ref{tab:ccs_for_h0} whilst the total with cross-matched ages from GAMA are shown in the second column. There are 161 correctly classified true CCs which we will call the ``correct'' subsample. This group represents the ideal scenario in which all galaxies that are identified as CCs with traditional methods are also correctly identified by the ML pipeline. Additionally, we compare two more subsamples that represent a more realistic scenario in which there are incorrectly classified CCs included in the sample of true CCs. For these two subsamples, we take the correct subsample and add the misclassified galaxies. Firstly, we create the ``missed'' subsample, which consists of the 161 correct, true CCs and the 35 true CCs that are misclassified as non-CCs by the K-means algorithm to create a subsample of 196 true CC galaxies. We then create the ``false'' subsample which also consists of the 161 correct, true CCs. However, we then add the 37 non-CC galaxies that are falsely classified as CCs by the K-means algorithm to create a third and final subsample of 198 galaxies. In Table~\ref{tab:ccs_for_h0} these samples are the ``correct'', true CC + missed (``missed'') and true CC + false (``false'') which represent a more strict selection using ML, a selection that is more indicative of how CCs are traditionally selected and a ``contaminated" sample. For the comparison of $H_0$ calculations we use these subsamples to compare how contamination affects our results.

\begin{figure}
    \centering
    \includegraphics[width=\linewidth]{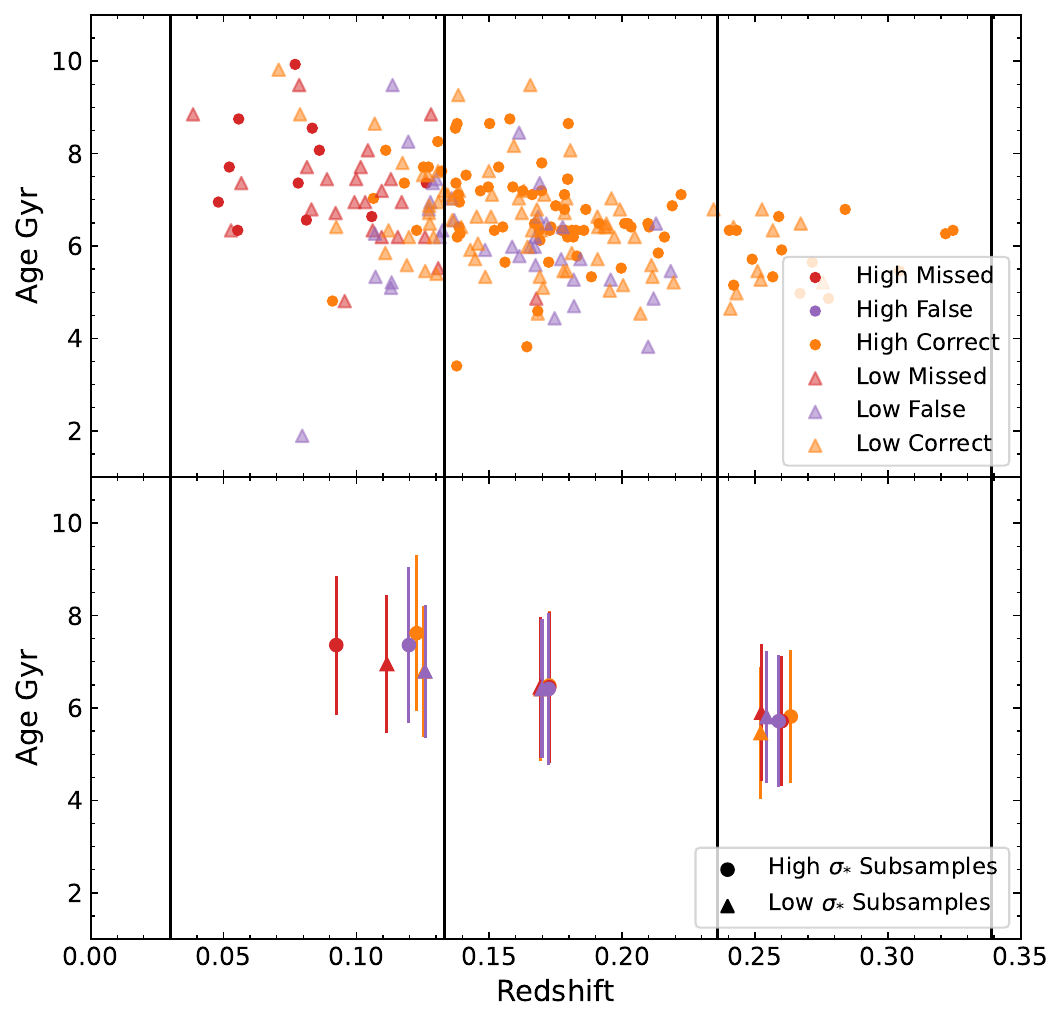}
    \caption{The \textit{top panel} shows the distribution of galaxy ages across redshift for each subsample with respect to the redshift bins at z=[0.03, 0.133, 0.236, 0.339] shown with solid black lines. The ``correct'', ``missed'' and ``false'' sample is shown in orange, red and purple respectively with the high $\sigma_*$ galaxies shown with fully opaque circles and the low $\sigma_*$ sample being shown with transparent triangles. In the \textit{bottom panel}, we show the median values for each subsample with the average age uncertainties of each bin.}
    \label{fig:cc_age_redshift}
\end{figure}

We split each of our subsamples in half using their median stellar velocity dispersions $\langle \sigma_*\rangle$ shown in Table~\ref{tab:cc_veldisp}. This effectively separates the most massive CCs with the highest velocity dispersions \citep{2022ApJ...928L...4B}. We then split each $\sigma_*$ subsample into three redshift bins each, in which the redshift range ($\Delta z\simeq 0.103$) corresponds to approximately the average age uncertainty ($\Delta$age~$\simeq 1.5$~Gyr). Fig.~\ref{fig:cc_age_redshift} shows the different subsamples.

\begin{table}
    \centering
    \caption{The different subsamples split by their respective median stellar velocity dispersions $\langle \sigma_* \rangle$.}
    \begin{tabular}{c|ccc}
        Subsample & $\langle\sigma_*\rangle$ kms$^{-1}$ & No. Low & No. High \\
        \hline
        Correct (161) & 273 & 80 & 81 \\
        Missed (196) & 271 & 99 & 99 \\
        False (198) & 267 & 98 & 98 \\
    \end{tabular}
    \label{tab:cc_veldisp}
\end{table}

\subsection{Constraining the Hubble Constant} \label{calc_h0}
The Hubble constant and the cosmological parameters it depends on can be constrained using the age-redshift relationships of cosmic chronometers by using the ages of the oldest objects at redshift intervals in order to place lower limits on the age of the Universe $t_U(z)$ \citep{2019JCAP...03..043J, 2021PhRvD.103j3509K, 2022JHEAp..36...27V, 2022ApJ...928L...4B}. By considering the time at which a galaxy forms $t_{\text{form}}$ after the Big Bang we may determine an age-redshift relationship:

\begin{equation} \label{eq:age-t_form}
    \text{age}(z) = t_U(z) - t_{\text{form}}
\end{equation} 

We follow the method derived by \cite{2022ApJ...928L...4B} to constrain $H_0$ using Markov Chain Monte Carlo sampling. This method predicts the most likely values for $H_0$, $\Omega_{m,0}$ and $t_{\text{form}}$ that satisfy the function:

\begin{equation} \label{eq:tu(z)} 
    t_U(z) = \frac{1}{H_0} \int_{z}^{\infty}{\frac{dz'}{(1+z')E(z')}}
\end{equation} 

where $E(z)$ is the normalised Hubble parameter: 

\begin{equation}
    E(z)=\sqrt{\Omega_{m,0}(1+z)^3+(1-\Omega_{m,0})f(z)}
\end{equation}

and

\begin{equation}
    f(z)=(1+z)^{3(1+w_0+w_a)}\exp^{-3w_a(z/(1+z))}
\end{equation}

This assumes that late-time expansion history is described by a flat $w_0w_a$ CDM Universe in which the dark energy equation of state varies with cosmic time under the CPL \citep{2001IJMPD..10..213C, 2003PhRvL..90i1301L} parameterisation such that $w_0=-1$ and $w_a=0$ for a flat $\Lambda$CDM Universe. Therefore, we fit eq.~\ref{eq:tu(z)} with $w_0=-1$ and $w_a=0$.

By considering the galaxy downsizing scenario, we can place additional constraints on the age-redshift relation by assuming a constant offset of the time of formation. This is because more massive galaxies form their stellar population at earlier times in comparison with less massive galaxies therefore by considering the high and low $\sigma_*$ populations we can rewrite eq.~\ref{eq:age-t_form} as:

\begin{equation}
    \text{age}_{\text{low }\sigma_*}=\text{age}_{\text{high }\sigma_*}-\Delta t_{\text{form}}
\end{equation}

where $\Delta t_{\text{form}}$ is offset in formation time. 

MCMC sampling is a method of sampling from a probability distribution in order to construct the most likely distribution. Monte Carlo sampling typically draws independent samples from the distribution but Markov Chain Monte Carlo sampling draws samples based on the existing sample. We use the \texttt{emcee} \citep{2013PASP..125..306F} MCMC sampler to model three parameters $\theta= (t_{\text{form}}, H_{0}, \Omega_{m,0})$ and assume a Gaussian $\chi^2$-likelihood function $\propto e^{-\chi^2/2}$ with uniform, non-informative priors: $H_{0}\sim \mathcal{U}(1,150)$~kms$^{-1}$Mpc$^{-1}$, $\Omega_{m,0}\sim \mathcal{U}(0.01,0.99)$ and, $t_{\text{form}}\sim \mathcal{U}(1,10)$ Gyr. 

\begin{figure}
    \centering
    \includegraphics[width=\linewidth]{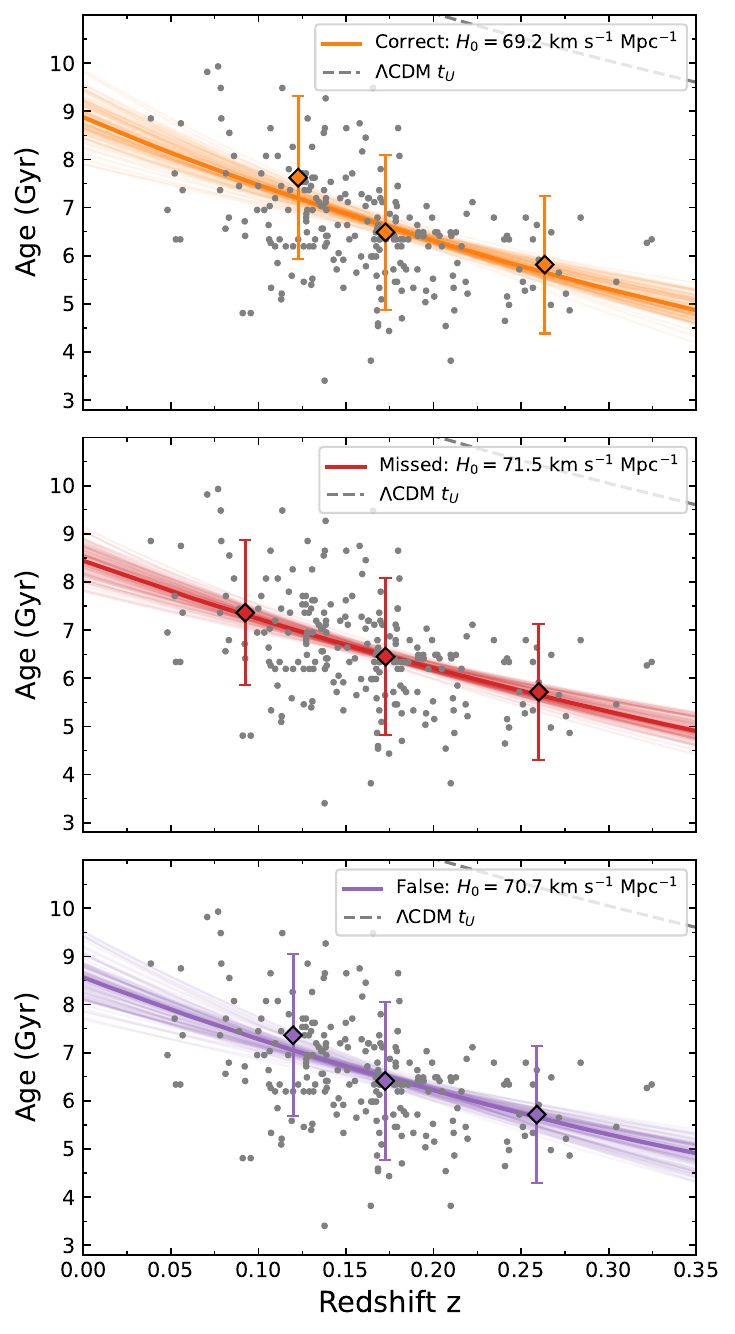}
    \caption[Illustrative function of age across redshift]{The age–redshift relation calculated using eq.~\ref{eq:tu(z)} for the correct, missed and false subsamples is shown in orange, red and purple, respectively. Faded curves represent 100 randomly selected draws from the joint posterior distributions, while solid curves show the median posterior predictions. Planck18 $\Lambda$CDM is plotted for illustrative purposes only with a grey dashed line. Diamonds represent the binned median observational ages with associated uncertainties.}
    \label{fig:h0_3bins}
\end{figure}

We initialise 32 walkers near the optimiser solution and run the chains for 10,000 steps, discarding the first 200 steps as burn-in and retaining every tenth subsequent sample. To illustrate how parameter uncertainties propagate into the predicted age-redshift relation, we plot 100 randomly selected realisations from the resulting joint posterior distribution, using all three parameters from each individual posterior draw in Fig.~\ref{fig:h0_3bins}.

We show the results for the three different subsamples in Table~\ref{tab:h0s} and plot each set of parameters using the $t_U(z)$ function in Fig.~\ref{fig:h0_3bins}. Our value of $H_0=71.46^{+28.41}_{-20.67}$~kms$^{-1}$Mpc$^{-1}$ using the missed subsample, which is the largest sample of CCs, is nearly the same as the reported value of $H_0=72^{+27}_{-19}$~kms$^{-1}$Mpc$^{-1}$ by \cite{2022ApJ...927..164B}. Our $\Omega_{m,0}$ and $t_{\text{form}}$ values are slightly higher but within the large uncertainties we share. 

\begin{table}
    \centering
    \begin{tabular}{c|ccc}
        Subsample & $H_0$~kms$^{-1}$Mpc$^{-1}$ & $\Omega_{m,0}$ & $t_\text{form}$~Gyr \\
        \hline
        Correct & $69.2^{+33.39}_{-20.29}$ & $0.404^{+0.38}_{-2.74}$ & $3.99^{+3.37}_{-2.2}$ \\
        \\
        Missed & $71.46^{+28.41}_{-20.67}$ & $0.43^{+0.378}_{-2.14}$ & $3.92^{+3.33}_{-2.14}$ \\
        \\
        False & $70.7^{+33.41}_{-21.65}$ & $0.42^{+0.36}_{-2.87}$ & $4.01^{+3.43}_{-2.21}$ \\
    \end{tabular}
    \caption[Results of MCMC sampling for $H_0$]{The results of the MCMC sampling for the high $\sigma_*$ subsample of each sample.}
    \label{tab:h0s}
\end{table}

\section{Discussion} \label{disc}
It is important to bear in mind that although the CC sample is chosen with strict selection cuts in order to minimise the presence of young stellar populations, ensuring these galaxies are amongst the oldest objects at their respective redshifts, there may be contamination as a result of issues with the spectroscopy such as aperture effects. For example, GAMA uses AAOmega spectroscopy using the Anglo-Australian Telescope \citep{2004SPIE.5492..410S} which has an aperture diameter of 2~arcseconds which means low redshift galaxies may not be fully covered by the fibre, therefore signatures from younger populations may be missed. Similarly, the BOSS spectrograph has a circular fibre of 3~arcseconds however VANDELS and LEGA-C are both observed using the VIsible Multi-Object Spectrograph (VIMOS) that has a rectangular slit with a fixed width of 1~arcsecond. VANDELS is configured for deep, faint, high-redshift galaxies and it can cover the entirety of galaxies between $1 < z < 1.5$ whereas LEGA-C will only cover the inner regions of galaxies between $0.6 < z < 1$. This means LEGA-C galaxies could have younger stellar populations that are not covered, therefore leading to more contamination of the final CC sample than in the VANDELS CC sample. Our CC candidates, observed with GAMA are the most affected by these aperture effects as they have relatively low redshifts with a small fibre diameter which means large portions of the outer regions of galaxies may be missed. This could be part of the reason why the ML pipeline sometimes misclassifies CCs and non-CCs that have intermediate properties. For example, galaxies that are labelled as CC candidates when classified using traditional methods but have been misclassified as non-CCs by the ML algorithm could indicate that only their inner region is free from young stellar populations rather than the galaxy as a whole. To this end, it would be useful to compare how different surveys hold up in terms of the ML classifications as surveys with spectroscopy that covers the entire galaxy may give us a more certain true set of CC candidates. This consideration is included as a caveat for the application of our CC selection and pipeline to galaxies observed by different surveys and at different redshifts. In particular, at lower redshift, the fibre may not encompass the full extent of a galaxy and could therefore exclude younger stellar populations that may contaminate the integrated galaxy properties. While this does not affect the photometry used in our analysis, it highlights a potential source of minor contamination when identifying CCs using spectroscopic measurements and should therefore be considered when applying our methodology to other datasets. A comparison of the ML classifications across surveys with different spatial coverage would be a useful avenue for future work, as spectroscopy that more completely samples the galaxy may provide a more representative assessment of whether a galaxy satisfies the CC criteria.

\subsection{Comparison of Model and Clustering Performance}
Our accuracy metrics are comparable, if not better, than similar studies based on ML based classification using autoencoding methods combined with clustering. \cite{2022RAA....22h5006H} compares the performance of a K-nearest neighbours (KNN) algorithm with a CAE+KNN model and an attention-based CAE and KNN algorithm which adds an attention module \citep[CBAM;][]{2019JEI....28b3012L} to allow the CAE to focus on what is meaningful in the galaxy images they train with. Their results are reported for different tests in which different morphologies are assigned to be outliers. The test that is most similar to our work is their fifth test that takes one class of completely smooth galaxies as their inlying sample and a number of classes to be the outliers including edge-on, spiral and cigar-shaped galaxies. Their tests are not directly comparable as our ``inlying'' sample is the equivalent of their outlying sample as it features all types of galaxies except for CCs whereas our outlying sample is all one type of CC. However, they report similar recall, accuracy and F1-scores for their CAE+KNN and attention CAE+KNN with typical values ranging between 0.56-0.91 depending on the test and model. Their best model has recall, accuracy and F1-scores of 0.78, 0.96 and 0.78 respectively which classifies edge-on galaxies as outliers and smooth round galaxies as inliers. Overall, they find their models struggle with similar broad shapes (e.g. round smooth galaxies and face-on spirals) but perform better at classifying round versus elongated which suggests their models are slightly biased towards the large-scale structure as opposed to finer detail. We find similar results when considering the misclassified galaxies in the false and missed subsamples. The non-CC misclassified as CCs appear to have more elliptical morphology whereas the true CCs misclassified as non-CCs have more elongated shapes that could be confused with lenticular or edge-on spirals. 

In a study using ML methods to classify isophotal profiles extracted from $grizY$ images from the DES survey, \cite{2022MNRAS.511.3330T} find that their XGBoost and PCA-based model correctly classifies 93\% of LTGs with 7\% being misclassified as ETGs. Whereas 14\% of ETGs are misclassified as LTGs with 86\% of ETGs being correctly classified. When we normalise our predictions for the $u-g$, $g-r$ and $u-$ band MSE model we find we correctly classify 81.4\% of non-CCs and misclassify 17.6\% of them as CCs whereas 82.4\% of CCs are correctly classified with 18.6\% misclassified as non-CCs. This is not a direct comparison as our CC and non-CC samples do not equate exactly to ETG and LTG as there are a number of ETGs in the non-CC group. However, it is interesting to note their model performs better at predicting late-type which is typically more complex. Our final model performs slightly better at classifying the CC sample which features more simple morphology however our test models (especially the colour-only models) all classify the non-CC galaxies better which feature a mix of LTG and ETG morphology.

\cite{2021A&A...648A.122V} compares a number of ML algorithms when classifying galaxy morphology using photometry from SDSS DR9. Part of their analysis uses feature importance to rank a number of photometric properties including ellipticity measurements and different colours for which they find the $u-r$ colour to have the highest importance. We find that using the $u-r$ colour in addition to the features used for our $u-g$, $g-r$ and MSE K-means model performs worse than our model. We then try $u-r$ with $g-r$ and MSE which results in matching performance with our model and finally $u-r$ with $u-g$ and MSE which performs the worst. A summary of these results are shown in table~\ref{tab:vav}. We conclude that the $u-r$ and $u-g$ colours perform functionally the same for our application. Additionally, the Random Forest feature importance results from \cite{2023A&A...671A..99E} shows $g-r$ is the highest ranking feature for selecting quiescent galaxies using various photometric colours with $g-i$ and $r-i$ closely following. Out of the SDSS colours that they test, $g-r$ ranks highly for the CatBoostClassifier and XGBClassifier feature importance too, with the $u-g$ and $u-r$ colours following. To summarise, the colours we find are the most important are the $u-g$, $g-r$ and $r-i$.

\begin{table*}
    \centering
    \begin{tabular}{c|cccc}
        Input Features & Accuracy & Precision & Recall & F1-Score \\
        \hline
        $(u-g)$ + $(g-r)$ + MSE & 0.819 & 0.816 & 0.824 & 0.82 \\
        $(u-g)$ + $(g-r)$ + $(u-r)$ + MSE & 0.798 & 0.991 & 0.603 & 0.75 \\
        $(u-g)$ + $(u-r)$ + MSE & 0.289 & 0.033 & 0.015 & 0.021 \\
        $(g-r)$ + $(u-r)$ + MSE & 0.819 & 0.816 & 0.824 & 0.82 \\
    \end{tabular}
    \caption[Comparison of different colour choices]{Comparison of different combinations of input features to evaluate the difference between the $u-g$, $g-r$ and $u-r$ colours. }
    \label{tab:vav}
\end{table*}

\section{Conclusions} \label{conc}
We present a pipeline to flag potential cosmic chronometer candidates using the combination of a convolutional autoencoder trained on the images of galaxies in tandem with a K-Means clustering classifier. The K-means classifier is able to identify cosmic chronometer candidates using the MSE reconstruction losses generated by the CAE trained with $u-$band images in combination with the $u-g$ and $g-r$ photometric colours with an accuracy of 0.819 and recall of 0.824. This is found by classifying 199 CCs and 199 non-CCs in which the final pipeline correctly classifies 164 and 162 respectively. We confirm the K-means model is robust to different splits of the data by calculating the AUROC and PR-AUC using repeated stratified K-fold cross-validation for which we report the AUROC~$=0.897\pm0.015 (95\%$~CI~$ 0.865-0.926)$ and PR-AUC~$=0.877\pm0.022(95\%$~CI~$0.834-0.918)$.

We compare the performance of the classifier with and without the CAE reconstruction losses and show that it performs best when provided with the MSE reconstruction loss calculated between the input and output $u-$band images. By using the CAE reconstruction loss we are therefore able to utilise anomaly detection as a new method of identifying CCs as anomalous galaxies within our sample. This allows us to employ machine learning techniques that would otherwise not be effective because of the issues relating to the CC dataset being imbalanced due to their intrinsic rarity. In addition to the $u-$band MSE reconstruction loss improving the performance of the classifier, we assess a number of combinations of colour magnitudes in the classifier and find that the ``bluer'' colours ($u-g$ and $g-r$) improve performance significantly in comparison with ``redder'' combinations or combinations with more or less colours (eg. $u-g$ alone or $u-g$, $g-r$ and $r-i$ together).

We compare the correctly classified galaxies with those that are misclassified by the pipeline (ie. 35 CCs misclassified as non-CC and 37 non-CCs misclassified as CC). When evaluating the performance of the classifier, we find that the true CCs misclassified as non-CC appear to have intermediate features as they fall in the overlap of the true CC and non-CC populations. This is clear when comparing the traditional selection criteria, redshift and stellar population properties. Additionally, when comparing the visual morphology in $u-$band images, the misclassified non-CCs appear more similar to the correctly classified CC sample and vice versa. This indicates the pipeline struggles with images that exhibit intermediate properties. 

Finally, we compare the calculated $H_0$ values when using the ML-identified CCs in order to evaluate the effect of contamination when using misclassified galaxies. We use three subsamples by combining the correctly classified CCs with the two subsamples of misclassified galaxies. These three subsamples encapsulate the most obvious (according to the K-means algorithm) CCs using the correct subsample, the more realistic case in which there are less obvious CCs using the missed subsample and a contaminated sample in which there are misclassified non-CCs combined with true CCs. When we calculate the Hubble constant using these samples we find, $H_0=69.2^{+33.39}_{-20.29}$~kms$^{-1}$Mpc$^{-1}$ for the ``correct'' subsample,  $H_0=71.46^{+28.41}_{-20.67}$~kms$^{-1}$Mpc$^{-1}$ for the ``missed'' subsample and $H_0=70.7^{+33.41}_{-21.65}$~kms$^{-1}$Mpc$^{-1}$ ``false'' for the false subsample. It is important to remember these values of $H_0$ are purely for comparing the effect of misclassifications and should not be taken as robust calculations as our samples of CCs are limited by high age uncertainties and a relatively shallow redshift range. However, they do show that when including the misclassified non-CCs, the uncertainty of $H_0$ is larger. This confirms that the CC identification pipeline should only be used as a method of flagging potential CCs for spectroscopic based confirmation using the traditional selection criteria rather than as a foolproof tool of identification. With the advent of large imaging surveys like LSST and Euclid this tool is vital in the search for identifying potential CCs and thus creating large datasets for $H_0$ calculation as imaging surveys that cover vast portions of the sky can identify likely CCs for confirmation through traditional, spectroscopy based criteria. LSST filters include $u-$band imaging which allows for direct application with some fine-tuning of the model at low-redshift whereas Euclid will provide wide-area imaging with its VIS and NISP instruments. The broad VIS passband overlaps restframe $u-$band wavelengths over approximately $0.5\lesssim z\lesssim1.5$, providing sensitivity to the restframe near-UV emission of CCs at higher redshift. Although recovering restframe $u-$band properties from VIS will require SED modelling rather than providing a direct $u-$band measurement, the combination of VIS and NISP photometry could enable the model to be extended to higher-redshift CCs. Retraining of the model with VIS and NISP is therefore an alternative now that we have established it is possible to identify CCs using imaging and photometry.

\section{Acknowledgements}
We would like to acknowledge the Viper High Performance Computing facility of the University of Hull and its support team. Additionally, the GAMA survey which is a joint European-Australasian project based around a spectroscopic campaign using the Anglo-Australian Telescope. The GAMA input catalogue is based on data taken from the Sloan Digital Sky Survey and the UKIRT Infrared Deep Sky Survey. Complementary imaging of the GAMA regions was obtained by a number of independent survey programmes including GALEX MIS, VST KiDS, VISTA VIKING, WISE, Herschel-ATLAS, GMRT and ASKAP providing UV to radio coverage. GAMA is funded by the STFC (UK), the ARC (Australia), the AAO, and the participating institutions. The GAMA website is https://www.gama-survey.org/ .

\section{Data Availability}
Our cosmic chronometer sample will be available upon publication, please see Paper I \citep{2026MNRAS.tmp.1629H} for a full description of the traditional selection method. Additionally, both the trained convolutional autoencoder and K-means clustering models will be available on GitHub upon publication.

\bibliographystyle{mnras}
\bibliography{bibliography}
\label{lastpage}
\end{document}